\documentclass[a4paper,twocolumn,prb,showkeys]{revtex4}
\usepackage{epsfig}
\usepackage{isotope}
\usepackage{multirow}
\usepackage{array}
\usepackage{tabularx}
\usepackage{color}
\usepackage{ulem}
\usepackage{amssymb}
\usepackage[version=3]{mhchem}
\usepackage{ae,aecompl,aeguill}
\usepackage{pslatex}
\usepackage{setspace}
\usepackage{graphicx}
\usepackage{subfigure}
\usepackage{lipsum}
\usepackage{verbatim}
\usepackage{longtable}
\usepackage{dcolumn}
\usepackage{bbold}
\usepackage{bbm}

\newcommand{\isoO}{\isotope[17]{O}}
\newcommand{\isoNa}{\isotope[23]{Na}}

\newcommand{\isoP}{\isotope[31]{P}}

\newcommand{\isoGa}{\isotope[69]{Ga}}
\newcommand{\isoAs}{\isotope[75]{As}}

\newcommand{\Q}{$\mathrm{Q}^1$}
\newcommand{\QQ}{$\mathrm{Q}^2$}
\newcommand{\QQQ}{$\mathrm{Q}^3$}
\newcommand{\Qn}{$\mathrm{Q}^n$}

\newcommand{\VV}{{ \textbf V} }
\newcommand{\UU}{{ \textbf U} }
 
\DeclareTextSymbol{\degre}{T1}{6}
\DeclareTextSymbol{\degre}{OT1}{23}

\graphicspath{{fig/}}

\begin{document}


\title{Extended Czjzek model applied to NMR parameter distributions in sodium metaphosphate glass}

\author{Filipe Vasconcelos}
\altaffiliation{Current address: CEA, IRAMIS, SIS2M, CEA/CNRS UMR 3299 - Laboratoire de Structure et Dynamique par R\'esonance Magn\'etique
F-91191 Gif-sur-Yvette cedex, France}
\author{Sylvain Cristol, Jean-Fran\c cois Paul,   Laurent Delevoye}
\affiliation{Unit\'e de Catalyse et Chimie du Solide, UMR CNRS 8181,
\'Ecole Nationale Sup\'erieure de Chimie de Lille, Universit\'e  de
 Lille, BP 108,   59652    Villeneuve d'Ascq    Cedex,    France}

\author{Francesco Mauri}
\affiliation{Institut de Min\'eralogie et Physique des Milieux
Condens\'es, Universit\'e Pierre et Marie Curie, Campus Boucicaut,
140 rue de Lourmel, 75015 Paris}

\author{Thibault Charpentier}
\affiliation{CEA, IRAMIS, SIS2M, CEA/CNRS UMR 3299 - Laboratoire de Structure et Dynamique par R\'esonance Magn\'etique
F-91191 Gif-sur-Yvette cedex, France}

\author{G\'erard Le Ca\"er}
\affiliation{Institut de Physique de Rennes, UMR UR1-CNRS 6251,
Universit\'e de Rennes 1, Campus de Beaulieu, B\^at 11A, Avenue du G\'en\'eral Leclerc,
F-35042 Rennes Cedex, France}

\date{\today}

\begin{abstract}
\noindent
The Extended Czjzek Model (ECM) is applied to the distribution 
of NMR parameters of a simple glass model (sodium metaphosphate, $\mathrm{NaPO_3}$) obtained by Molecular Dynamics (MD) simulations. 
Accurate NMR tensors, Electric Field Gradient (EFG) and Chemical Shift Anisotropy (CSA), are calculated  
from Density Functional Theory (DFT) within the well-established PAW/GIPAW 
framework. Theoretical results are compared to experimental high-resolution solid-state NMR data and are used 
to validate the considered structural model. 
The distributions of the calculated coupling 
constant $C_Q\propto |V_{zz}|$ and of the asymmetry parameter $\eta_Q$ that characterize the 
quadrupolar interaction are  discussed in terms of structural considerations 
with the help of a simple point charge model. Finally, the ECM analysis is  shown 
to be relevant for studying  the distribution of CSA tensor parameters and
gives new insight into the structural characterization of disordered systems by solid-state NMR.
\end{abstract}

\maketitle

\newpage

                                          \section{Introduction}

It     is    generally     accepted,    since    the early     work    of
Zachariasen,~\cite{JACS.54.3841}  that oxide  glasses are  built  up from
polyhedra randomly organised
in such a way however that the local short-range order that prevails in the corresponding crystalline compounds is preserved. 
This seminal model, named the Continuous Random Network (CRN), evolved slightly with
time and was extended to covalent glasses such as amorphous silicon (Refs.~\onlinecite{gibson2012solving,treacy2012local} and
references therein). Its validity appears to be reinforced by diffraction techniques
from which radial distribution function (RDF) can be determined and compared to the
radial distribution associated with the local arrangements of polyhedra in a CRN model.
However, conclusions about the structures of amorphous materials are questionable when
they are obtained solely from pair correlations as they suffer from a lack of uniqueness
beyond very short range structure~\cite{gibson2012solving}. Other techniques that are more sensitive to
topological or medium-range order are needed to constrain structural models as advocated
very recently for amorphous silicon~\cite{gibson2012solving,treacy2012local}.
Complementary techniques are then needed to provide
additional input data which constrain models determined for instance by Reverse Monte Carlo
methods. Among the various spectroscopic techniques that can be used (Raman, Infrared, EXAFS for instance),
Solid State Nuclear Magnetic Resonance (NMR) spectroscopy can
be seen as one of the most promising tools to characterise oxide glass structure.

In solid-state NMR, the  structural information is carried by the different interactions, 
i.e. chemical shift, quadrupolar, dipolar coupling, indirect  spin-spin coupling, 
which are deduced either directly by a one-dimensional experiment or indirectly through 
the use of multiple-dimensional experiments (homo and hetero-nuclear correlations). 
These interactions being orientationally dependent, they often contribute 
to a line-broadening of NMR spectra.  
All these interactions can be described by tensors, i.e., chemical shift anisotropy tensor
(CSA), Electric Field Gradient tensor (EFG). If the anisotropic parts of NMR interactions 
have been for a long time considered as a major drawback for the NMR investigation of solids, 
the development of Magic Angle Spinning (MAS) technique has open new perspectives by significantly 
improving the spectral resolution, especially for those nuclei with a spin half value ($I=1/2$). 
For nuclei with higher spin values ($I>1/2$), which are subjected to the quadrupole 
interaction, MAS technique alone is unable to average out completely the quadrupolar
interaction. Over the last twenty years, several methods were proposed to get rid of the 
second-order  anisotropic quadrupolar interactions: Double Orientation Rotation (DOR),~\cite{MP.65.1013} 
Dynamic-Angle Spinning (DAS),~\cite{CPL.152.248} Multiple-Quantum-MAS (MQMAS),~\cite{JACS.117.5367,JACS.117.12779} 
or satellite-transition MAS (STMAS).~\cite{JACS.122.3242} 
The most versatile and most widely used MQMAS method is a high-resolution experiment designed for 
half-integer quadrupolar nuclei and routinely used to average out the anisotropic 
second-order quadrupolar interactions by correlating the multiple-quantum transitions 
with the single quantum transitions. These  methods, among others, allow nowadays the observation 
of many nuclei from the periodic table and the characterisation of their anisotropic 
interaction tensors.

In crystalline solids, the broadening of NMR spectra of powdered samples is 
the result of different orientations of the crystallites relative to the external field.  
In such a case, the structural characterisation can be done without ambiguity from the principal 
components of the tensors alone.
In the case of an amorphous system, 
the distribution of chemical environments results in an intrinsic distribution of all
the components of the interaction tensor.
This additional distribution is then responsible for the broadening of 
the spectra observed experimentally in glasses. 
The challenging problem in solid-state NMR analysis of disordered 
materials is to interpret this spectral broadening in structural terms.
First, the extraction of NMR parameter distributions from NMR data can be rather difficult. It is clear
that the availability of analytical models can facilitate this task. Second, 
the relationhips between NMR parameter distributions and the structural and chemical disorder
has to be established.

Structure elucidation by solid-state NMR suffers from a fundamental 
drawback: \textit{the assignment of NMR spectra to chemical 
environments is an indirect process}. Thereby, except for some 
cases where the chemical environments belong clearly to very different families, 
the assignment needs a structural model to remove possible 
ambiguities.
The development of methods to calculate 
NMR parameters from the atomic-scale during the last decade is this view as they allow 
the assignment of similar chemical environments and confirm the 
sensibility of NMR parameters to structure. 
In particular, the now routinely used DFT-PAW/GIPAW (Projector Augmented Wave and Gauge Including Projector Augmented Wave respectively)~\cite{PRB.63.245101,JACS.125.541}
combined approach accounts very accurately for the CSA and EFG tensors of a large amount of crystalline 
organic~\cite{JACS.127.10216,JACS.129.8932} or inorganic
~\cite{JACS.125.541,JPCB.108.13249,JACS.126.12628,JPCB.109.7245,JACS.129.13213} compounds.
However, this theoretical approach to assign resonances is 
still an indirect process. Indeed, for crystalline compounds, 
a known structure, deduced from diffraction techniques, 
is mandatory to be used as an input in \textit{ab initio} and 
DFT codes and to further be compared to experimental NMR results. 
Similarly, for amorphous or disordered compounds, structural model 
must be built to confirm the interpretation of NMR spectra. 
Recently, the combination of molecular dynamics (MD) and solid-state 
NMR has shown a great ability to interpret experimental spectra and 
to propose structural glass models for different oxide glasses
~\cite{JPCB.108.4147,JPCB.109.6052,JPCC.113.7917,PRL.101.065504,SSS.12.183,JPCM.22.145501,PCCP.12.6054}. 
This approach has the advantage to give access to the distributions
of all the NMR interaction tensors (CSA, EFG). 

The case of EFG tensor distribution has been recently subject to 
some new considerations in the context of solid-state NMR. 
Indeed, D'Espinose de Lacaillerie \textit{et al.}~\cite{JMR.192.244} 
examined the fields of applications of the Czjzek model~\cite{PRB.23.2513,JPCM.10.10715}, 
also called Gaussian Isotropic Model (GIM) that is
 summarized in the background section (Section~\ref{sec:background}). 
The GIM is the first analytical model for the analysis of 
NMR lineshapes observed in the MAS NMR spectra of quadrupolar 
nuclei in disordered solids. Apart from its application in other 
spectroscopies (ex. in M\"ossbauer spectroscopy Ref.~\onlinecite{JPCM.10.10715,PRB.28.4944} and references therein), 
this model was successfully applied to the NMR study of 
amorphous systems, to different nuclei such as 
aluminium-27~\cite{GCA.68.5071,JMR.192.244,JPCB.114.1775}, 
gallium-71~\cite{SSNMR.14.181,JPCM.12.5775,PCCP.12.11517,IC.50.8252}, 
arsenic-75~\cite{PCCP.12.11517} and should be applicable to other nuclei 
(sodium shows characteristic isotropic distributions see ex. 
Fig. 2 of Ref.~\onlinecite{PCCP.7.2384}). In particular, this model 
is able to describe the asymmetric broadening observed at low 
chemical shift~\cite{JMR.192.244,IC.50.8252}. Since a quadrupolar 
nucleus in an amorphous system presents also a distribution of 
the isotropic chemical shift, the GIM gives the means to better 
separate and quantify these two contributions within the resonance broadening. 
However, the specific two hypotheses defining the GIM (i.e., 
rotational invariance and central limit theorem)~\cite{JPCM.10.10715,JPCM.22.065402} 
are unable to provide structural considerations from the analysis 
of the NMR lineshape distribution. As discussed by 
Le Ca\"er \textit{et al.},~\cite{JPCM.22.065402} 
the GIM, which results \textit{in fine} from the application of a central limit theorem  to the EFG tensor (section~\ref{subsec:gim}), 
can thus be seen as a kind of ``black hole'' 
from \textit{which no information about the specific structural features of the investigated 
solid and about the physical origins of the EFG can come}. 

The introduction of the Extended Czjzek Model (ECM),~\cite{JPhysColl.46.LC,JPCM.10.10715,JPCM.22.065402} 
also summarized in Section~\ref{sec:background}, is currently the simplest but useful way to generalise the GIM. 
In a recent contribution, one of us showed that this ECM could be seen as the introduction 
of physical (i.e., structural) contribution to the GIM. Indeed, the ECM 
\textit{is intended to mimic the EFG contribution of a well-defined neighbourhood of a 
given atomic species modified by the effect of more remote atomic shells.} 
In practice, this well-defined first atomic shell and the remote atomic shells 
are modelled by a fixed ``local'' contribution and a distributed Czjzek contribution, 
respectively. Therefore, depending on the relative weight of the local contribution, 
the ECM is able to include some direct structural effects in the Czjzek model. Indeed, for a given EFG tensor, the relation 
between structural data and the quadrupolar interaction can be easily determined from 
\textit{ab initio} calculations of the EFG~\cite{JMR.192.244}. Then, the 
observation of an ECM distribution with a predominant local contribution can be 
interpreted through structural considerations.

In this paper, we present a general approach to study the EFG distribution of a simple glass model, 
namely the sodium metaphosphate glass (\ce{NaPO3}), generated by molecular dynamics (MD). In a previous work, the NMR signature 
of the non-bridging oxygen (NBO) of this compound was interpreted, to some extent, with a 
continuous random network model~\cite{IC.47.7327}. Indeed, the observed small distribution 
of the quadrupolar parameters, the mean value of which is comparable to those observed in related
crystalline compounds, was interpreted as the conservation of the local structure.  
In addition, the large distribution of the isotropic chemical shift was interpreted 
as the signature of long-range disorder. For bridging oxygens (BO), some correlation 
between the quadrupolar parameters and simple structural properties appeared to be close 
to those observed in silicates.~\cite{PRB.70.064202} This glass system can thus be seen as 
a good candidate for the application of the ECM to \isoO~environments. In a second step, 
the use of the ECM to describe sodium NMR lineshapes will show the ability of such model to 
reveal a pure isotropic tensor distribution as defined in the GIM.    

This paper is organised as follows.  In section~\ref{sec:background}, we 
recall the basic characteristics of the distributions of the main EFG parameters in the frame of the GIM and ECM.
In Section~\ref{sec:Method}, we present the methodology used to generate different 
configurations of the sodium metaphosphate glass and to calculate the NMR parameters for all 
nuclei in the corresponding glass structure. In section~\ref{sec:Results}, the results are compared 
to experimental \isoO~ and \isoP~high-resolved NMR spectra to validate the structural model. 
In the same section, we present a detailed analysis of the calculated EFG distribution within the ECM. 
In Section~\ref{sec:STRUCTanalysis}, we analyse the distributions of the EFG tensor parameters and their link with the structure of the glass. 
In particular, we show that a simple point charge model analysis is able to explain 
the distributions related to the EFG tensor for the different sites. In Section~\ref{sec:CSA}, 
we introduce a preliminary application of the ECM to the distribution of the Chemical Shielding Anisotropy tensor. 

\begin{figure*}
\centering
\includegraphics[width=0.8\textwidth]{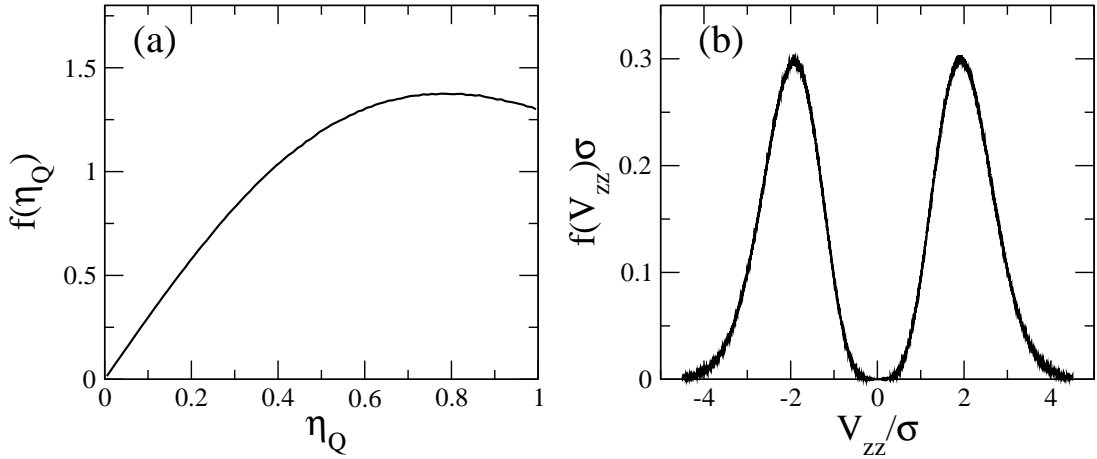}
\caption{Characteristic shapes of the marginal distributions of (a)
$\eta_Q$ and (b) $V_{zz}/\sigma$ in case of a Czjzek distribution
of the EFG tensor (GIM). (a) The distribution $f(\eta_Q)$ is independent of $\sigma$,
the single free parameter of the bivariate distribution of $V_{zz}$ and $\eta_Q$;
the mean asymmetry parameter is ~0.6098 (b) The $V_{zz}$ distribution is symmetric about zero.
The ratio $\rho_Z$ (see Eq.~\ref{eq:rhoz}) is independent of sigma, being $\sim$~0.32607.
\label{fig:czjzek}}
\end{figure*}

                         \section{Distribution of the EFG tensor: background}
\label{sec:background}

\subsection{EFG related definitions}
\label{subsec:def}
In solid-state NMR, the  quadrupolar interaction is  characterised by two parameters: 
$C_Q$ and $\eta_Q$ (respectively the quadrupolar coupling constant, and  the 
quadrupolar asymmetry), which are both defined from the  three principal  components  
of the diagonal  EFG tensor  ($\bf{V}$), by the following relations: 
\begin{equation}
\label{eq:quad}
C_Q=\frac{e|Q|}{h}|V_{zz}| ,\;\;\;\;\;\; \eta_Q=\frac{V_{yy}-V_{xx}}{V_{zz}}
\end{equation}
where $e$ is the elementary charge, $h$ is the Planck's constant, $Q$
is the quadrupolar  moment and the principal components $V_{ii},\ (i=x,y,z)$ are sorted 
such that  $|V_{zz}|  \geqslant |V_{xx}| \geqslant |V_{yy}|$. In the present work, 
we used $Q=-25.58 \times 10^{-31} m^2$ and $Q=100.54\times 10^{-31} m^2$, 
for respectively the oxygen and the sodium quadrupolar moment as tabulated by Pyykk\"o~\cite{MP.99.1617}. 

Thereafter, we only consider the parameters $V_{zz}$ and $\eta_Q$ to represent the principal 
components of the diagonal EFG tensor. Even if NMR experiments at room temperature are unable 
to provide the sign of the largest principal component~\cite{B-Abragam}, it is explicitly available 
by DFT calculations and is important to characterize the distribution as demonstrated below.

The EFG tensor, whose components are $v_{i,j}$ with $(i,j=x,y,z)$, is usually not diagonal,  and three others parameters defining the orientation of the tensor in a fixed reference frame are required.  
Consequently, the EFG tensor is completely defined by five independent 
quantities as expected for a traceless symmetric second-rank tensor. 
Following Czjzek \textit{et al.}~\cite{PRB.23.2513}, we define five real parameters 
$U_i$ from the Cartesian components of the EFG tensor.  
\begin{equation}
\label{eq:U_i}
U_1=v_{zz}/2, U_2=\frac{v_{xz}}{\sqrt{3}},U_3=\frac{v_{yz}}{\sqrt{3}},U_4=\frac{v_{xy}}{\sqrt{3}},U_5=\frac{(v_{xx}-v_{yy})}{2\sqrt{3}}  
\end{equation}
The five-dimensional vector ${\bf{U}}=(U_1,\ldots,U_5)$ constitutes a random vector 
representative of the EFG tensor of an amorphous solid in the same fixed reference frame 
for all the sites. 
The distribution of $\UU$ fulfils a number of conditions 
for disordered solids which are statistically invariant by any rotation. This 
invariance does not imply any kind of local geometrical symmetry. In other words, 
``statistical isotropy'', which is a global property, and ``geometrical anisotropy'', 
which is a local property, are not contradictory characteristics and are most often 
the rules for amorphous solids. Statistical isotropy simply means that the 
distribution of $\UU$ remains unchanged when any fixed, but arbitrary, frame of 
reference is chosen to calculate the EFG's of all atoms of the selected isotope. 
It implies nothing about possible symmetries of local clusters centered on these 
atoms. The essential conditions, which are needed in the discussion of the results 
of the present article, are derived by a simple method in Ref.~\onlinecite{JPCM.10.10715}.
They are:
\begin{itemize}
\item The distribution of $\UU$ is such that $\langle U_i\rangle= $0 and 
$\langle U_iU_j\rangle=\sigma^2\delta_{ij}$ with $(i,j=1,\ldots,5)$ where $\sigma^2$ is the common variance of the five components of 
$\UU$. The latter conditions are true for any distribution of $\UU$ as soon as means and variances do exist.
\item The marginal distribution $P(U_1)$ is a priori asymmetric and different 
from the marginal distributions $P(U_{k,k>1})$ which are all identical and symmetric.
\end{itemize}
When structural models are available, theoretical distributions $P(U_i)$ can be used to check 
if the previous conditions of statistical isotropy hold or not. In addition, when the distribution 
of $\UU$ is multivariate Gaussian, as in the Czjzek model, then the $U_i$ are independent random 
variables and the five distributions $P(U_i)$ are identical Gaussians.

\subsection{Gaussian Isotropic Model (Czjzek model)}
\label{subsec:gim}
The Czjzek model~\cite{PRB.23.2513}
is used for the analysis of the joint distribution of the components of the previous vector
$P(U_1,\ldots,U_5)$ in the context of an isotropic distribution of the EFG tensor. 
At the thermodynamic limit (i.e. an infinite number of sites), 
or at least when the physics that determine the distribution of the EFG 
tensor meets the quite general requirements of the multidimensional central 
limit theorem, the random variables $U_i$ become independent and identically 
distributed according to a Gauss distribution with a zero mean.
With these two assumptions, the Czjzek model is summarised by the following analytical 
expression for the bivariate distribution $P(V_{zz}, \eta_Q)$:
\begin{equation}
\label{eq:distrib_czjzek}
P(V_{zz},\eta_Q)=\frac{1}{(2\pi)^{1/2}\sigma^5}V_{zz}^4\eta_Q(1-\eta_Q^2/9)\exp\Big(-\frac{S^2}{2\sigma^2}\Big)
\end{equation}
where $S$ is the norm of the tensor ($S^2=V_{zz}^2(1+\frac{\eta_Q^{2}}{3})$). 
This model is fully defined by a single parameter, namely the standard deviation $\sigma$ 
of the Gaussian distribution of every component $U_i$. 
In this paper, we only consider some 
properties of the marginal distributions $f(\eta_Q)$  and  $f(V_{zz})$. 
For further description of the GIM properties, the reader is referred to the following 
articles (Refs.~\onlinecite{JMR.192.244,PRB.23.2513,JPCM.10.10715} and~\onlinecite{JPCM.22.065402}).

Figure~\ref{fig:czjzek} presents the characteristic shapes of the
marginal distributions $f(\eta_Q)$  and  $f(V_{zz})$ obtained from
the distribution~(\ref{eq:distrib_czjzek}). These distributions have the
following properties:
\begin{itemize}
\item[\textbf{(A)}] the shape and the mean value of the $f(\eta_Q)$ distribution are independent 
of $\sigma$ (Figure~\ref{fig:czjzek} (a)). The latter distribution is given by
\begin{equation}
\label{eq:dis_eta_czjzek}
f(\eta_Q)=3\eta_Q \frac{1-\frac{\eta_Q^2}{9}}{\Big(1+\frac{\eta_Q^2}{3}\Big)^{5/2}}
\end{equation}
\item[\textbf{(B)}] The marginal distribution $f(V_{zz})$ is symmetric about zero (Figure~\ref{fig:czjzek} (b))
and  $f(V_{zz})\propto~V_{zz}^{4}$ for small values of $V_{zz}$.
\end{itemize}

\begin{figure*}
\centering
\includegraphics[width=0.8\textwidth]{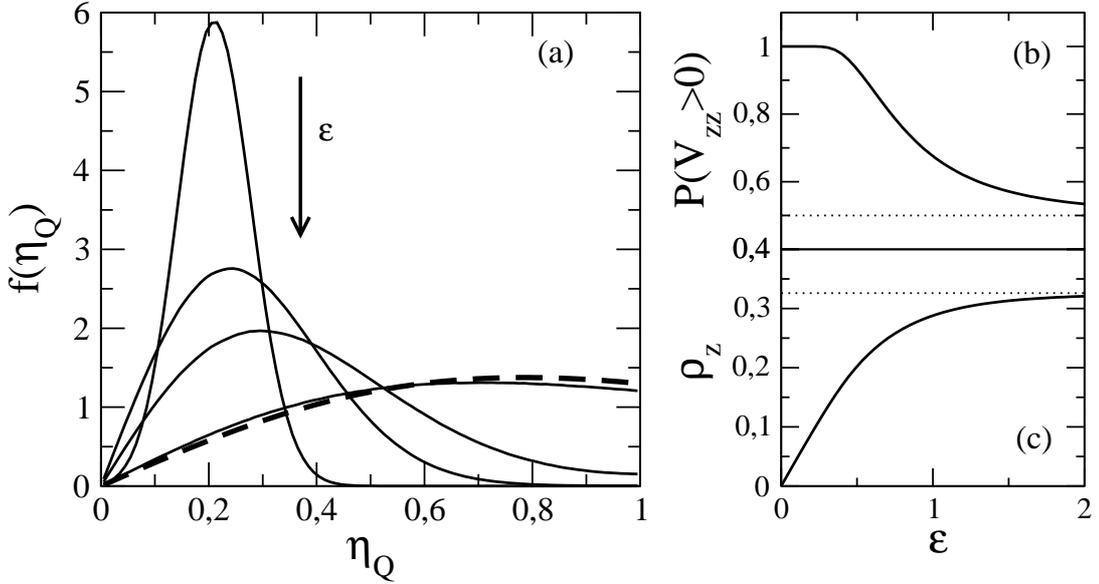}
\caption{Evolution of (a) the shape of the $\eta_Q$ distribution,
(b) the probability of occurrence of $V_{zz}>0$ and (c) the parameter
$\rho_Z$ (relation~\ref{eq:rhoz}) as a function of the parameter $\epsilon$
of the ECM. The other ECM parameters are: $\eta_Q(0)$=0.2, $V_{zz}(0)$=1. 
The dashed lines represents (a) the $\eta_Q$ distribution and (b) the asymptotical values of $P(V_{zz}>0)$ and (c) $\rho_Z$ for a Czjzek model. 
\label{fig:nq_vzz_rho}}
\end{figure*}

\subsection{Extended Czjzek model}

\subsubsection{Model presentation}
\label{sec:ecm}
The Extended Czjzek Model (ECM) allows the introduction of an anisotropic part in the total EFG tensor. 
In this model, the total observed EFG tensor $\bf{V}(\epsilon)$ 
is defined as the sum of two distinct contributions as follows: 
\begin{equation}
\label{eq:czjzek_etendue}
\bf{V}(\epsilon)=\bf{V_0}+\rho\bf{V_{\mathrm{GIM}}}
\end{equation}
where $\bf{V_0}$ represents the local anisotropic tensor
due to a close neigbourhood of the considered atom and $\bf{V_{\mathrm{GIM}}}$ is the global isotropic 
distribution modelled by a GIM tensor weighted by a parameter $\rho$ due to more remote atomic shells.
The choice of a Czjzek contribution to express the effect of noise is justified by the two very general
assumptions, statistical isotropy and gaussianity, which unavoidably lead to the Czjek model.
In Eq.~\ref{eq:czjzek_etendue} the tensor $\bf{V_{\mathrm{GIM}}}$ is obtained 
from a vector $\UU$ (Eq.~\ref{eq:U_i}) whose components are standard Gaussians 
with means equal to zero and variances equal to 1.
All tensors in Eq.~\ref{eq:czjzek_etendue} are expressed without loss of generality in the local 
frame of reference in which $\VV_0$ is diagonal (see \ref{sec:appendix1} and~\ref{sec:appendix2} ).  
Thus, the distribution of $\bf{V}(\epsilon)$ is not statistically isotropic. 
If needed , $\bf{V}(\epsilon)$ can be transformed into a tensor $\VV'(\epsilon)$ which is statistically isotropic (\ref{sec:appendix1} ). 
The tensors $\VV(\epsilon)$ and $\VV'(\epsilon)$ have identical distributions of principal values (\ref{sec:appendix2}). 
The distributions of the components of $\VV'_0$ are further discussed in \ref{sec:appendix2}.

The total EFG tensor is now a function of $\epsilon$, with $\rho$ defined by:
\begin{equation}
\label{eq:epsilon}
\rho=\frac{\epsilon||\bf{V_{0}}||}{||\bf{V_{\mathrm{GIM}}}||}
\end{equation}
which corresponds to the ratio of the norm of the different contributions. 
This $\epsilon$ parameter permits to study the influence of each contribution 
independently of the value of the fixed tensor. 
The  principal values of the fixed diagonal tensor $\VV_0$ are fully characterized by 
$V_{zz}(0)$ and $\eta_Q(0)$.
In summary, the ECM is defined by three parameters: $V_{zz}(0)$ and $\eta_Q(0)$ for the 
local contribution and $\epsilon$ for the weight of the noise in the total EFG tensor.
Hereafter, $\VV_0$ will be consistently named the ``local''
contribution to the EFG tensor $\VV(\epsilon)$ while the name ``noise''
(or Czjzek or background ) 
will be used to designate the second contribution.

The main simplification of the ECM is to consider that $\eta_Q(0)$ and $V_{zz}(0)$ are not distributed, being
the same for all atoms of a given family. By family, we mean the set of all local
clusters, centered on atoms of a given species, which can be put in coincidence
by some rotation, within very small atomic displacements (see sections 3 of
Ref.~\onlinecite{JPCM.10.10715} and 3.1 of Ref.~\onlinecite{JPCM.22.065402} )
This simplification aims at restricting the number of free parameters to a
minimum value while mimicking the essential structural effects of the local atomic configurations. A
disordered solid, for which the ECM is relevant, does not necessarily include a single family of sites.
Many families may be necessary to describe the whole set of sites occupied by a given atomic
species. A finite collection of extended Czjzek families may be sufficient to describe the EFG
properties of some solids while others need to be described by a continuous distribution of families ( Appendix~\ref{subsec:perturbed_diamond} ) .

\subsubsection{$\eta_Q$ and $V_{zz}$ distributions within the ECM}
\label{subsubsec:ECMprocedure}

We restrict the following discussion to the characteristics of the distributions of $\eta_Q$ and $V_{zz}$ in the ECM model and we compare them to the related distributions in the Czjzek model as summarized by the previous points \textbf{(A)} and \textbf{(B)}.
For a detailed description of all the ECM properties, 
the reader is referred to Ref.~\onlinecite{JPCM.22.065402}.

Figure~\ref{fig:nq_vzz_rho} (a) presents the evolution of the distribution $f(\eta_Q)$ of $\eta_Q$ with $\epsilon$ for $\eta_Q(0)=$ 0.2. 
For small values of $\epsilon$ (i.e. large local contribution), 
the distribution of $\eta_Q$ is narrow and concentrated around $\eta_Q(0)$.
For increasing $\epsilon$ (i.e. increase of the global isotropic contribution) the distribution $f(\eta_Q)$
converges to the $\eta_Q$ distribution of the Czjzek model (Eq.~\ref{eq:dis_eta_czjzek}).
For all pairs ($\eta_Q(0),\epsilon$), the $f(\eta_Q)$ distribution is accurately approximated
by a closed-form expression (Eq. 22 of Ref.~\onlinecite{JPCM.22.065402}). 
For a given value of $\epsilon$, the $\eta_Q$ distributions allow a determination of the 
parameter $\eta_Q(0)$ of the ECM.

Figure~\ref{fig:nq_vzz_rho} (b,c) presents the evolution with $\epsilon$ of two characteritics of the $V_{zz}$ distributions.
Figure~\ref{fig:nq_vzz_rho} (b) presents the variation with $\epsilon$ of the probability of occurence 
of positive values of $V_{zz}$ ( i.e $P(V_{zz}>0)$).
A given probability $p=max(P(V_{zz}>0), P(V_{zz}<0))$, with $0.5\leq p\leq  1$ is not in a one-to-one correspondence with $\epsilon$ 
(see for instance Fig. 6 of Ref.~\onlinecite{JPCM.22.065402} ). 
It is thus necessary to select another parameter for an unambiguous determination 
of $\epsilon$. 
A convenient parameter, which characterizes the $V_{zz}$ distribution 
while being scale independent, denoted $\rho_Z$, is defined as follows:
\begin{equation}
\label{eq:rhoz}
 \rho_Z=\frac{\sigma(|V_{zz}|)}{\langle|V_{zz}|\rangle}
\end{equation}
where $\sigma(|V_{zz}|)$ and $\langle|V_{zz}|\rangle$ are respectively the standard deviation and
the average value of $|V_{zz}|$ distribution.
Figure~\ref{fig:nq_vzz_rho} (c) presents the evolution of $\rho_Z$ with $\epsilon$. 
The ratio $\rho_Z$ is accurately approximated the relation given in (i) below.
The previous ratio allows an unequivocal determination of $\epsilon$. 

We propose now a simple procedure to analyse, within the ECM, a given distribution of 
$V_{zz}$ and $\eta_Q$. Indeed, as on the one hand the distribution of $V_{zz}$ 
is practically independent of $\eta_Q(0)$ and depends largely on the $V_{zz}(0)$ 
value and on the other hand $\epsilon$ is completely defined 
from the $V_{zz}$ distribution, only three steps are needed to determine all the ECM parameters:

\begin{itemize}

\item[(i)]  $\epsilon$ is determined from $\rho_Z$ using the expression (Eq. 20 of Ref.~\onlinecite{JPCM.22.065402}) 
$$\rho_Z=0.32607(1-\exp(-2.097\epsilon^{1.151}))$$ valid for any $\eta_Q(0)$, where 0.32607 is the value for the Czjzek model.

\item[(ii)] $\eta_Q(0)$ is first determined from the $\eta_Q$ distribution using the approximation (Eq. 22 of Ref.~\onlinecite{JPCM.22.065402})
$$f(\eta_Q) \propto \eta_Q^{\alpha}\exp(-k\eta_Q^{\beta})+(2-\eta_Q)^{\alpha}\exp(-k(2-\eta_Q)^{\beta})$$
with $0\leq\eta_Q\leq1$ and where $k$, $\alpha$, $\beta$ are tabulated in Ref.~\onlinecite{JPCM.22.065402} as a function of $\epsilon$.

\item[(iii)] Using the previous parameters, $|V_{zz}(0)|$ is finally well approximated by:
$$|V_{zz}(0)| \propto \frac{\langle|V_{zz}|\rangle \sqrt{1+ \langle\eta_Q\rangle^{2}/3}}{\tau(\epsilon)\sqrt{1+\eta_Q(0)^{2}/3}}$$ where $\tau(\epsilon)$ is a function given by Eq.15 of Ref.~\onlinecite{JPCM.22.065402}
\end{itemize}

The method described above involves only simple desk calculations. 
We notice that the deconvolution program of solid-state quadrupolar NMR spectra of Grimminck et al.\cite{Grimminck2011114} 
uses a more precise method based on the bivariate distribution of the ECM expressed in the form of a triple integral.  

\begin{figure}
\centering
\includegraphics[width=0.45\textwidth]{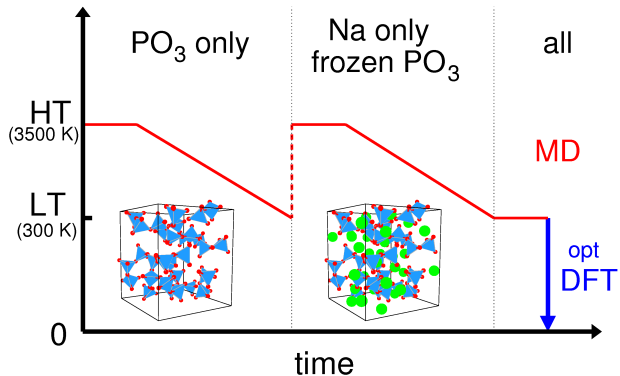}
\caption{(Color on-line) Schematic representation of the procedure used to obtain one
configuration of sodium metaphosphate glass. From a well equilibrated trajectory
at high temperature, the phosphate network is first quenched to a low temperature configuration.
Then, the same quench rate is applied to the sodium atoms in the frozen phosphate ``skeleton''.
After a small equilibration trajectory at low temperature on all atoms, the DFT total energy of
the configuration is minimised at 0 K. A total of 17 configurations
were generated by this procedure.
\label{fig:quench_procedure}}
\end{figure}

                             \section{Theoretical approach}
\label{sec:Method}

\subsection{Sodium metaphosphate structural glass model}

In the present work, we used a combination of classical molecular dynamics (MD) 
simulations and DFT calculations to generate a configuration manifold of the sodium metaphosphate 
glass system as schematized in Figure~\ref{fig:quench_procedure}.

Basically, our approach is divided into two general steps: (i) the quench from a 
high-temperature configuration to a low temperature configuration using 
classical MD simulations and (ii) the energy minimisation of the final configuration 
within DFT (which can be seen as a quench to 0 K with infinite rate). 
In other words, the first step allows a quick sampling of different configuration 
space basins by fast classical MD. The second step allows to sample the 
``DFT local minima'' of the potential energy landscape~\cite{NATURE.410.259}. 
This latter step is essential to obtain accurate local geometries required to derive 
reliable NMR parameters \cite{JPCB.108.13249}. 
However, as the size of the system is limited to about a hundred of atoms 
due to DFT-PAW/GIPAW calculation time, more than one glass configuration is required
to correctly sample the different NMR tensor distributions. Following the 
ergodic hypothesis, one can generate several configurations to accurately sample 
the configuration space to overcome this size limitation~\cite{PRB.71.024208}. 
Configurations extracted from low-temperature trajectories 
cannot be used in this aim as at low temperature, the structural reorganisations 
are too small and NMR parameters are limited to vibrational time averaging~\cite{JACS.132.5993}. 
Therefore, several quenches are needed. 

As classical molecular dynamic simulations are based on empirical potentials, 
it is important to carefully check its ability to generate reliable structures before DFT optimisation. 
In case of oxide glasses, 
an important structural feature is the coordination of the different polyhedra 
commonly represented within the \Qn~notation. For phosphates, $n$ is the number of 
phosphate linked to a given phosphate group with $n\leq$4. Thereafter, we call 
``secondary structure'' this coordination structure property, opposed to 
``primary structure'' (distances) which can be determined by 
radial distribution function analysis. This secondary structure 
is hardly modified at the \textit{ab initio} step of our procedure, 
even if we consider an AIMD at low temperature~\cite{JPCB.110.25810}. 
Indeed, the time scale currently available by AIMD is too short to induce such 
structural rearrangement. It is thus essential to correctly reproduce 
it at the classical MD level. 

Experimentaly it is known from \isoP~NMR that the binary sodium metaphosphate glass 
(\ce{(Na2O)1/2})\ce{(P2O5)1/2}) is only composed by \QQ~(e.g. infinite phosphates chains)~\cite{JNCS.263.1}.
It is also known that the \ce{Na2O} oxide play the role of network modifier.
Indeed, an excess or a default of \ce{Na2O} oxide in sodium metaphosphate glasses arises 
to respectively an increase of \Q~or \QQQ~coordination~\cite{JNCS.311.223}.

In a preliminary study, we observed that the classical force-field chosen for this study
(presented in next section) was unable to reproduce the experimental phosphate network 
of sodium  metaphosphate, and this for any reachable order of magnitude of the quenching rate. 
This behaviour is largely dependent on the sodium charge involved in the long range 
electrostatic term. Indeed, we observed that decreasing the sodium charge leads to an increase 
of the \QQ~concentration. Surprisingly, the \QQ~concentration reaches the experimentaly expected concentration 
when no sodium atoms were added at all. Consequently, we separated 
the classical molecular dynamics procedure into two steps, quenching first the phosphate 
network to obtain the correct \Qn~concentration and then quenching 
the sodium cations in the fixed phosphate ``skeleton''.

\subsection{Force-Field and Thermal quenching procedure}
Amorphous sodium metaphosphate configurations were generated using the effective 
force field proposed by van Beest, Kramer and van Santen (BKS)~\cite{PRL.64.1955,PRB.43.5068}. 
This force-field, which is only based on two-body potential, is formalised by the following equation:
\begin{equation}
\label{eq:force_field}
V_{\alpha\beta} = \frac{q_{\alpha}q_{\beta}}{r} + A_{\alpha\beta} exp(-\frac{r}{\rho_{\alpha\beta}}) - \frac{C_{\alpha\beta}}{r^6}
\end{equation}
where the first term is the electrostatic interaction with partial charges $q_{\alpha}$.  
The following two terms correspond to the Buckingham potential, which is composed by an 
exponential repulsive part and an $r^{-6}$ attractive term. 
Table~\ref{tab:BKS_forcefield} summarizes the values of the set of parameters ($q_{\alpha}$, $A_{\alpha\beta}$, 
$\rho_{\alpha\beta}$ and $C_{\alpha\beta}$) used in the classical step of our procedure for each 
$\alpha,\beta$ atomic pairs. The short-range part of the 
potential was truncated at 6.1\AA~and Ewald sum method was used to deal with the long-range 
electrostatic forces. 
The form A of \ce{NaPO3} crystalline compound~\cite{AC.B24.1621} 
was chosen as an initial configuration for the MD simulation. We used a 2$\times$2$\times$1 
 super-cell containing 160 atoms (32~\isoNa , 32~\isoP~ and 96 \isoO~ sites) with cell parameters set to $a=$ 12.20 \AA, $b=$ 12.48 \AA~and $c=$ 14.07 \AA~to match 
to the experimental density of the glass system with the same stoichiometry (2.53 g/ml). 

The configurations are obtained 
by quenching different configurations extracted from a long 
well equilibrated high temperature trajectory ($\sim$1 ns at 3500K) 
to low temperature configurations (300K). The classical 
quench was obtained by 20 temperature steps in NVE ensemble 
(equilibrating for 2 ps at each step). The former quench 
procedure corresponds to a quenching rate of $\sim$4 K/ps. 
The chosen quench rate is slow enough to reproduce final structures
that provide NMR parameters in reasonable agreement with experiments and fast
enough to be tractable in terms of computation time.
Classical trajectories  were obtained using the academic 
code DL\_POLY~\cite{DL_POLY}. 
Hundreds of configurations have been obtained by this procedure. 
Then, the atomic positions were optimised (e.g energy minimisation) with DFT calculations
using the VASP code (Vienna Ab-initio Simulation Package)~\cite{CompMatSci.6.15,PRB.54.11169,PRB.49.14251,PRB.48.13115}
These calculations were done at DFT-GGA 
(PW91)~\cite{PRB.54.16533,PRB.46.6671,PRB.48.4978} level of theory
 using standard PAW~\cite{PRB.59.1758} to describe the electron-ion interactions. 
Plane-wave basis cut-off was set to $\sim$ 44 Ry (600eV) 
and only the $\Gamma$-point was used in the Brillouin integration. 
The geometry was considered as converged when the forces acting on the atoms were below 0.01 eV/\AA. 
This last step of the procedure being more time consuming, only a 
limited number of configurations could be generated. 
We finally generated 17 configurations of metaphosphate glass (\ce{NaPO3}) by 
this complete procedure giving us a statistical sampling of 544 
sodium and  phosphorus sites and 1632 oxygen sites.   

\begin{table}[!t]
\begin{ruledtabular}
\begin{tabular}{ccccc}
\bf{Atomic pair} & \multicolumn{4}{c}{\bf{Force-Field parameters}} \\
$\alpha-\beta$ & $A_{\alpha\beta}$ &$\rho_{\alpha\beta}$  &
$C_{\alpha\beta}$ & $q_{\alpha}$ (charge) \\
\hline
\ce{Na-O} & 354.22072 & 0.24186 & 0.0 & $q_{\rm Na}$ (+1) \\
\ce{P-O}  & 903.4208 & 0.19264 & 1.98793 & $q_{\rm P}$ (+3.4)\\
\ce{O-O}  & 138.8773 & 0.36232 & 17.50 & $q_{\rm O}$ (-1.2)\\
\end{tabular}
\end{ruledtabular}
\caption{Force Field parameters used to generate MD trajectories in the classical step of the procedure.
This force-field was proposed by van Beest, Kramer and van Santen (BKS)~\cite{PRL.64.1955,PRB.43.5068}
Units: energies are given in J/mol , distances in \AA~and charges in atomic units.
\label{tab:BKS_forcefield}}
\end{table}


\begin{table}[!t]
\begin{ruledtabular}
\begin{tabular}{cccc}
& {\bf MD\footnotemark[1] } & {\bf Neutron\footnotemark[2] } & {\bf{this study}}  \\
\hline
Atomic pair                       &            R (\AA)          &              R (\AA)                 &      R (\AA)       \\
\ce{P-NBO}                        &            1.50             &                1.48                  &       1.45         \\
\ce{P-BO}                         &            1.59             &                1.61                  &       1.65         \\
\ce{P-P}                          &            3.18             &                2.93                  &       2.95         \\
\ce{O-O}                          &            2.51             &                2.52                  &       2.55         \\
\ce{Na-Na}                        &            3.10             &                3.07                  &       3.35         \\
\ce{Na-O}                         &            2.31             &                2.33                  &       2.35         \\
\hline
Coordination(\%)                  &                             &                                      &                    \\
\Q                                &            25               &                 -                    &       4            \\
\QQ                               &            50               &               100\footnotemark[3]    &      92            \\
\QQQ                              &            25               &                 -                    &       4            \\
\end{tabular}
\footnotetext[1]{from Ref.~\onlinecite{PCCP.1.173}}
\footnotetext[2]{from Ref.~\onlinecite{JPCM.19.415116}}
\footnotetext[3]{by definition of the metaphosphate structure}
\end{ruledtabular}
\caption{
A comparison between some structural properties of our MD configurations and published data.
(top) Position ($R$) of the first peak of the
radial distibution function. (bottom) Phosphate coordination (\Qn) composition.
Distances $R$ are given with an uncertainty of $\pm$ 0.02\AA\/ for the neutron study~\cite{JPCM.19.415116}
and of $\pm$ 0.025 for this work.
\label{tab:structure_local_comp}}
\end{table}


\subsection{NMR tensor calculations}
\label{subsec:calc}
NMR tensor calculations were performed on the obtained configurations 
using the PARATEC code at DFT level of theory~\cite{PARATEC,PR.140.1133}. 
We used the PBE \cite{PRL.77.3865} functional for the generalised gradient
approximation (GGA) of the exchange-correlation functional. 
The potentials due to the ions are represented by norm-conserving Troullier-Martins 
pseudo-potentials~\cite{PRB.43.1993}. The electronic configuration involved 
in the construction of the pseudo-potentials for the different nuclei \isotope[23]{Na},
\isotope[31]{P} and\isotope[17]{O} are respectively \{$2p^2$ $2p^6$ $3d^0$\},
\{$3s^2$ $2p^{1.8}$ $3d^{0.2}$\} and \{$2s^2$ $2p^3$\} with respective
core radii (in atomic unit) \{1.8 1.49 1.8\}, \{2.0 2.0 2.0\} and \{1.45 1.45\}

The electronic structure gives access to the EFG tensor through the reconstruction of the all-electron wavefunction
that is obtained with the PAW approach.~\cite{PRB.50.17953,JACS.125.541}
The non-diagonalised EFG tensor $\textbf{V}$ is used to calculate the 
five real components $U_i$ (Eq.~\ref{eq:U_i}), whereas the principal components $V_{xx}$, $V_{yy}$ 
and $V_{zz}$ of the diagonal tensor are used to calculate the quadrupolar 
coupling constant $C_Q$ and the asymmetry parameter $\eta_Q$ (defined by relation~\ref{eq:quad}). 
However, as mentioned above, $V_{zz}$ will be used in the distribution analysis.

The calculation of the chemical shielding tensor was performed using
the linear-response method \cite{PRL.77.5300} using the GIPAW 
reconstruction developed by Pickard and Mauri~\cite{PRB.63.245101}. 
Experimental isotropic chemical shifts $\delta_{cs}$ and absolute isotropic chemical
shielding $\sigma$ are related through the definition of an isotropic reference
shielding $\sigma_{ref}$ defined by $\delta_{cs}=\sigma_{ref}-\sigma$.
In the NMR community, it is commonly accepted to use a liquid  
as an external chemical shift reference. In the present work, we set the 
absolute chemical shift value to an unambiguous isolated resonance of a crystalline 
site studied previously~\cite{IC.47.7327}. 

For both EFG and CSA tensor calculations, we used an energy cutoff of 100 Ry, 
according to convergence tests previously performed on crystalline 
compounds~\cite{IC.47.7327,MRC.48.S142}. Owing to the size of the unit cell, only one k-point was used for the integration of reciprocal space.

\begin{figure}[!t]
\centering
\includegraphics[width=0.45\textwidth]{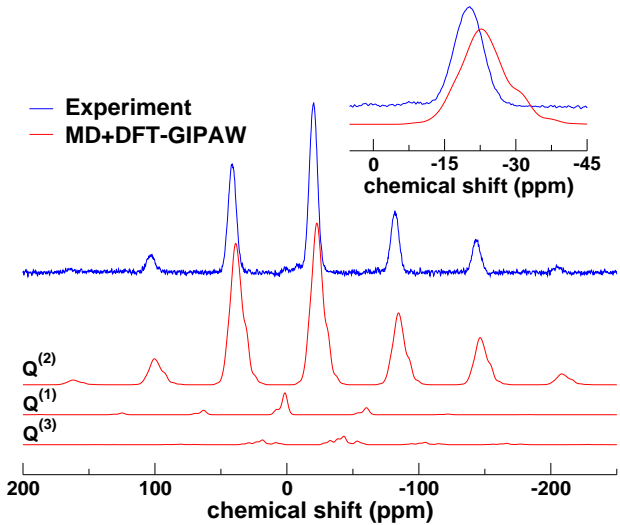}
\caption{(Color on-line) Experimental and simulated \isoP~MAS spectrum of \ce{NaPO3} at 18.8 T (MAS frequency is 20kHz). 
The simulated spectrum is obtained from DFT-GIPAW calculations on the MD configurations. 
The experimental spectrum shows small signals at 0 ppm assigned to \Q~sites revealing an excess in \ce{Na2O}. 
(insert) Enlarged view of the isotropic region of the spectrum centered on the \QQ~region.
\label{fig:31P_MAS} }
\end{figure}

                                           \section{Results}
\label{sec:Results}

\subsection{Validation of the structural model}

\subsubsection{Local order and medium range order}

Table~\ref{tab:structure_local_comp} compares the short (i.e the position of the maximum of the first peak of 
the radial distribution function, $R$) 
and medium range order parameters (relative concentration of~\Qn) of the 
\ce{NaPO3} MD configurations obtained with the procedure described above. 
We also present a previous attempt by Speghini \textit{et al.}~\cite{PCCP.1.173} 
to reproduce sodium metaphosphate glass structure from classical molecular 
dynamics. Experimental neutron diffraction data from Pickup \textit{et al.}~\cite{JPCM.19.415116} 
are also included for comparison. First, our glass model is seen to reproduce  
accurately the local structure, as represented through simple local geometrical 
parameters such as $R$. Our structural parameters are in better agreement with 
the experimental ones than are those of Speghini \textit{et al.} which were calculated 
only with a classical procedure. This is essentially due to the ab initio step in our calculation scheme.
However, as discussed above, an accurate structural glass model for phosphorous 
compounds cannot be restricted to the sole aim of reproducing the local structure. It should 
also describe the medium range structure. Our procedure provides a way to 
keep the ratio of \Q~ and \QQQ~at a low level to fit the experimental data, 
whereas for example, the relative concentration 
of \Qn~, is not accurately reproduced in the work of Speghini \textit{et al.} (see Table~\ref{tab:structure_local_comp}).

\begin{figure}[!t]
\centering
\includegraphics[width=0.45\textwidth]{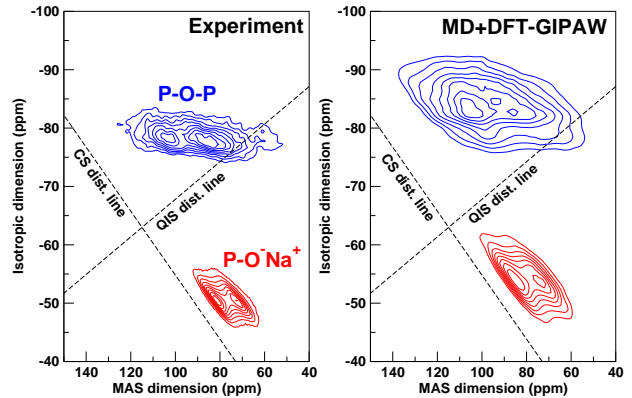}
\caption{(Color on-line) (left) Experimental and (right) theoretical~\isoO~MQMAS (calculated using MD configurations).
BO and NBO oxygens are respectively represented in blue (top contour plots) and red (bottom contour plots). 
Chemical Shift (CS) and Quadrupolar Isotropic Shift (QIS) lines are represented by dashed lines.
\label{fig:17O_MQMAS}}
\end{figure}

\subsubsection{Validation through \isoP~and \isoO~NMR}

Figure~\ref{fig:31P_MAS} exhibits the experimental \isotope[31]{P} 
MAS NMR spectrum of \ce{NaPO3} glass together with the one obtained from 
DFT-PAW/GIPAW calculation of NMR chemical shifts. The latter was 
obtained with 544 phosphorous sites. As expected, the main contribution
occurs around -20 ppm, a chemical shift usually assigned to \QQ~groups.
The experimental spectrum shows some very small resonances in the 
region of \Q~phosphorous, revealing a small excess of sodium in the chemical composition as 
compared to the stoechiometric composition. On the simulated spectrum, the small resonances
around 0 ppm and -40 ppm are due respectively to the \Q~and \QQQ~coordinations
present in our statistical sampling. A close look at the simulated 
spectrum additionally reveals an overestimation of the 
\isotope[31]{P}~\QQ~linewidth as  well as a small shoulder around
-30 ppm, revealing some limitations of our structural model.

Figure~\ref{fig:17O_MQMAS} compares experimental and simulated MQMAS
spectra of \ce{NaPO3} at 18.8T. It is important to recall that a 
MQMAS spectrum of an amorphous material gives the means to separate
the chemical shift distribution from the distribution of quadrupolar
parameters. Indeed, these distributions lead to a broadening of the
two-dimensional site along two different axis, as evidenced in 
figure~\ref{fig:17O_MQMAS} (a). In the NBO region, we interpreted recently
the broad distribution of chemical shift as a consequence
of long-range disorder in contrast to the conservation of the 
local structure observed through the small spreading along the quadrupolar
distribution axis.~\cite{IC.47.7327} The NBO and BO sites are 
experimentally clearly separated by their chemical shifts, quadrupolar
coupling constants and to a lesser extend by their asymmetry parameters. 

The NBO sites present a broad distribution in chemical shifts, which is
well reproduced by the MD configurations that permit to reproduce the resonance asymmetry observed
in the MAS dimension. 
On the other hand, for the BO, the line broadening in the MD spectrum reveals a broad
distribution of chemical shift which seems to be overestimated with
respect to the one observed experimentally. 
Overall, the main NMR parameters are well accounted for by the MD configurations
for both BO and NBO sites.
\begin{figure}[!t]
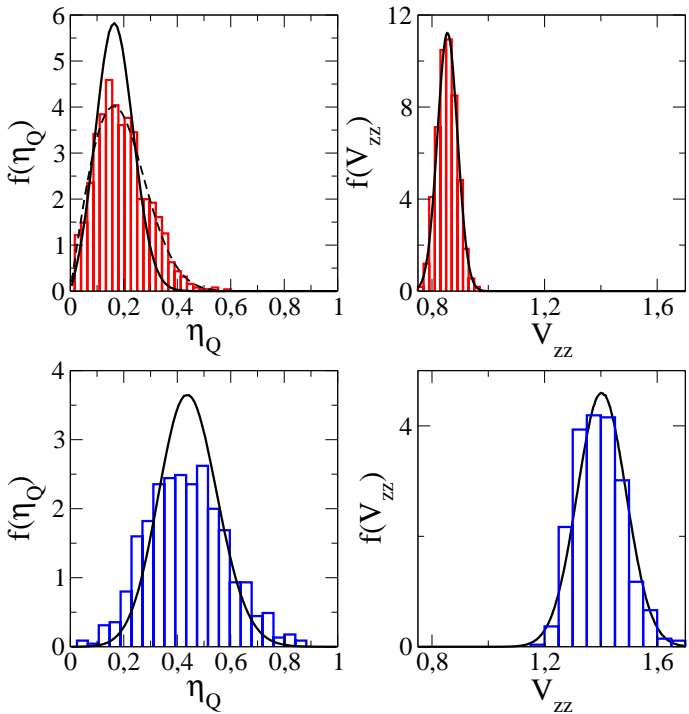

\centering
\includegraphics[width=0.5\textwidth]{ecm_17O_1}
\includegraphics[width=0.5\textwidth]{ecm_17O_2}
\caption{(Color on-line)
Distributions of $\eta_Q$ and $V_{zz}$ of \isoO~NBO (top and red) and of \isoO~BO (bottom and blue) for
the MD configurations compared with the $\eta_Q$ and $V_{zz}$ distributions obtained
from the ECM (black solid lines). The $\eta_Q$ distribution obtained when distributing
the ECM parameter $\eta_Q(0)$ according to a narrow Gaussian distribution ($\sigma^2=$ 0.006) is represented by the dashed line.
The other ECM parameters are: \isoO~NBO : $\eta_Q(0)=$ 0.15, $V_{zz}(0)=$ 0.853,
$\epsilon=$ 0.093.\isoO~BO: $\eta_Q(0)=$ 0.43, $V_{zz}(0)=$ 1.387, $\epsilon=$ 0.135.
\label{fig:BO_NBO_fit}}
\end{figure}

\begin{table*}[!t]
\begin{ruledtabular}
\begin{tabular*}{0.75\textwidth}{@{\extracolsep{\fill}}lcccccc}
& \multicolumn{3}{c}{\bf{MD Statistics}} & \multicolumn{3}{c}{\bf{ECM parameters}} \\
& N sites & $P(V_{zz}>0)$ & $\rho_z$         & $\epsilon$ & $V_{zz}(0)$ & $\eta_Q(0)$      \\
\hline
\isotope[23]{Na}    &   544   & $\sim$46\%   & 0.349           & -         &  0.148    &  -               \\
\isotope[17]{O} NBO &  1088   & 100\%        & 0.041           & 0.093     &  0.853    &  0.15            \\
\isotope[17]{O} BO  &   544   & 100\%        & 0.061           & 0.135     &  1.387    &  0.43            \\
\end{tabular*}
\caption{Statistical properties of the MD
configurations together with the ECM parameters obtained from the EFG
distribution analysis. (atomic units for EFG)
\label{tab:epsilon_deter}}
\end{ruledtabular}
\end{table*}

\subsection{ECM analysis of the EFG distribution}
\label{sec:ECManalysis}

Table~\ref{tab:epsilon_deter} gathers results obtained from the previous 
MD calculations and the ECM parameters deduced from 
the procedure described in section~\ref{subsubsec:ECMprocedure}.

Sodium atoms in our MD configurations have 
$P(V_{zz}>0) \approx P(V_{zz}<0)$ as expected if the GIM holds. 
This suggests that the EFG distribution of these sites is a 
Czjzek distribution. However the ratio $\rho_z$ (0.3487) is slightly larger than the 
Czjzek value of 0.32607. This difference is probably non-significant and related to  
the statistical fluctuations because of the low number of sites used to calculate it.
In any case, the ECM analysis of sodium sites leads to the same parameters than those of 
the Czjzek model.

The calculated EFG tensors of the oxygen atoms show a clear difference
between BO and NBO through the calculated ECM parameters, $V_{zz}(0)$ and 
$\eta_Q(0)$, as expected from the MD structures. The important 
element here is that oxygen sites present only positive values of $V_{zz}$,
differing strongly from the essentially equal weights of the positive and 
negative parts of the $V_{zz}$ distribution which holds for the GIM.
Furthermore, the $\epsilon$ values which are found to be small, being 0.09 and 0.13 
for the NBO and BO, respectively, reflect a large local contribution in the ECM. 
Figure~\ref{fig:BO_NBO_fit} presents the MD distributions of $\eta_Q$ 
and $V_{zz}$ parameters compared to the corresponding ECM 
distributions calculated with the parameters given in Table~\ref{tab:epsilon_deter}. 
Small deviations between MD 
distributions and ECM analyses are observed for $\eta_Q$. This point 
will be addressed specifically in the next section. 
Finally, it is worth noting that such an overall agreement is 
obtained with a model based on three parameters only (i.e $V_{zz}(0)$, 
$\eta_Q(0)$ and $\epsilon$).

\begin{figure}[!t]
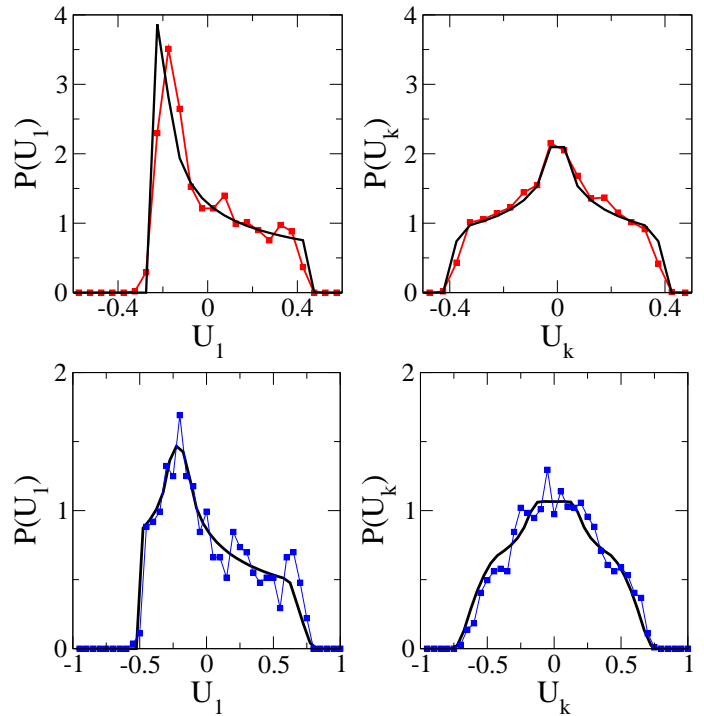

\centering
\includegraphics[width=0.5\textwidth]{onecharge_Ui}
\includegraphics[width=0.5\textwidth]{twocharge_Ui}
\caption{(Color on-line) Distributions of $U_1$ and  $U_{k,k>1}$
calculated for (top and red) \isoO~NBO and (bottom and blue) \isoO~BO from the MD configurations
compared respectively to the "one point charge model" and to the "two point charge model" (black solid lines).
The theoretical distributions are binned in the same way as the experimental ones. 
The parameters used to obtain the theoretical distributions (\ref{subsec:append_onecharge} 
and~\ref{subsec:append_Uniformrot}) of NBO and BO sites are respectively 
$\beta_0=V_{zz}(0)/2=$ 0.425 , 0.695 and $\eta_Q(0)=$ 0.15, 0.43 .
\label{fig:Ui_EFG}}
\end{figure}

\begin{figure}[!b]
\centering
\includegraphics[width=0.45\textwidth]{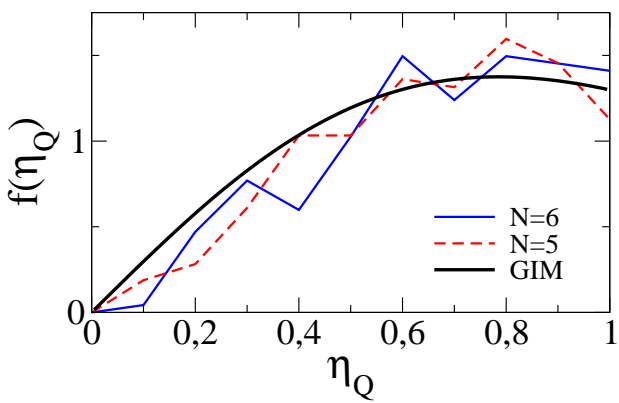}
\caption{(Color on-line) Distributions of $\eta_Q$ of \isoNa~in MD configurations
for two different coordination numbers ($N=$ 5,6).
These distributions are similar in shape to the Czjzek $\eta_Q$ distribution (solid line).
A similar trend is found for the other coordination numbers ($N=$ 4 and 7) despite the limited number of atoms used to calculate them.
($\sim$~50 sites).
\label{fig:coord_etaQ}}
\end{figure}

                     \section{Structural interpretation of the distributions}
\label{sec:STRUCTanalysis}
The previous discussion concentrates on the distributions of the principal values 
of the EFG as calculated from our structural models. 
In this section, we focus on the distributions of the components of 
the EFG vector $\UU$ (Eq.~\ref{eq:U_i}). 
These distributions are characterized and compared with some simple model predictions 
for the oxygen and sodium sites.

\subsection{The case of oxygen (\isotope[17]{O})}
\label{subsec:STRUCTanalysis_17O}
The calculated EFG distribution can be explained in terms of the 
local structural features of BO and NBO sites.
Indeed, the ECM analysis unveils the existence of a 
strong local contribution, consistent with 
the small coordination number of oxygen atoms. 
Intuitively, we understand
that the electrostatic field felt locally by oxygen atoms is principally
due to the variation of charge density induced by the chemical bonds shared with phosphorus.
All other contributions can be considered as small perturbations to this
``localised'' anisotropy in the charge distribution. 
As a consequence, it makes sense to analyse the
calculated~\isoO~EFG distribution with a simple model of effective point charges, 
as discussed for instance in Ref.~\onlinecite{JPCM.10.10715}.
This model describes the probability distribution of the EFG created 
at the centre of a sphere by a random repartition of $n$ point charges
on its surface. 
For this purpose, we analyse the distribution of $\UU$, the vector whose components 
are related to the elements of the EFG tensor through Eq.~\ref{eq:U_i}.
In practice, the discrete charge model 
is equivalent to the calculation of the five components of $\UU$ by 
the following classical relations:
\begin{eqnarray}
U_1 &=&\frac{1}{2}        \sum_n   (q_n/r_n^3)(3\cos^2(\theta_n)-1) \nonumber\\
U_2 &=&\sqrt{3}           \sum_n   (q_n/r_n^3)\sin(\theta_n)\cos(\theta_n)\cos(\phi_n) \nonumber\\
U_3 &=&\sqrt{3}           \sum_n   (q_n/r_n^3)\sin(\theta_n)\cos(\theta_n)\sin(\phi_n) 
\label{eq:allU}\\
U_4 &=&\frac{\sqrt{3}}{2} \sum_n   (q_n/r_n^3)\sin^2(\theta_n)\sin(2\phi_n) \nonumber\\
U_5 &=&\frac{\sqrt{3}}{2} \sum_n   (q_n/r_n^3)\sin^2(\theta_n)\cos(2\phi_n) \nonumber
\end{eqnarray}
where $q_n$  and $r_n,\theta_n,\phi_n$ are respectively the charge 
and the positions in spherical coordinates of the $n$th point charge. 

Le Ca\"er and Brand used this simple problem to exemplify  the convergence 
of the  distributions of the $U_i$'s $(i=1,\ldots,5)$  to the GIM distributions, 
that is to identical Gaussian distributions with variance $\sigma^2$, when the number 
$n$ of point charges increases. Numerical simulations were performed for $n=$ 2,3,4 
while the convergence to Gaussian distributions was proven when $n$ goes to infinity.
Actually, the five distributions converge rapidly to Gaussians which are excellent approximations for $n$
larger than 4-5. The EFG distribution is then the one predicted by the Czjzek model.
As required by the statistical invariance  by rotation of the previous point charge 
problem (section~\ref{subsec:def} and Ref.~\onlinecite{JPCM.10.10715}), the distribution of $U_1$ 
is found to be asymmetric while the distributions of the $U_{k,k>1}$ are identical 
and symmetric. 

Figure~\ref{fig:Ui_EFG} gives the distributions of the $U_i$ components
($U_1$ and $U_{k,k>1}$) for the~\isoO~sites (BO and NBO) encountered in 
our \ce{NaPO3} glass model. These distributions are compared to the discrete 
charge model described above. Using reduced units ($q_n=r_n=$ 1), a unique 
scaling factor is needed to match the two distributions. 
The~\isoO~sites are analysed with one or 
two charges, in order to mimic either the single covalent bond of NBO 
or the two covalent bonds of BO. 

The distribution obtained from a single charge accounts accurately for 
the distributions of the five $U_i$ of NBO. A first conclusion is that these distributions 
are fully consistent with statitical isotropy of the EFG tensor as described in 
section~\ref{subsec:def}. Second, these distributions are not very sensitive to a distribution 
of the scaling factor mentioned above. Indeed, Gaussian or uniform distributions 
of this factor do not change the central parts of the $P(U_k)$ distributions 
(figure~\ref{fig:Ui_EFG} top, see further ~\ref{subsec:append_onecharge} and figure~\ref{fig:U1_Uk_theo_12}). 
Further,  $P(U_1)$ present a sharp peak on the left side as suggested by the distribution obtained 
from the MD-DFT calculations (figure~\ref{fig:Ui_EFG}). 
It is important to note that even if the marginal distributions $P(U_k)$ are 
correctly described by the previous model, the distribution of $\eta_Q$ due to 
the effect of a single charge is a delta peak at $\eta_Q=$0 in contradiction with 
the calculated one (figure~\ref{fig:BO_NBO_fit}). 
Indeed, the axial symmetry of the single point charge yields an asymmetry parameter equal to zero. 
Therefore, it is 
necessary to add a background to the EFG of this  charge. Chosing a Czjzek EFG 
tensor to account for  this background and expressing the total EFG tensor in 
the frame of reference of the EFG due to the single charge yields the ECM 
(Eq.~\ref{eq:czjzek_etendue}). It is still an ECM in the presence of fluctuating 
scaling factors (see~\ref{sec:appendix2}).
Moreover, the Czjzek background leaves essentially unchanged the $U_i$ distributions. 
The latter have a rather limited discriminating power and are less suited than the distribution 
of $\eta_Q$ to evidence fine effects.
Slight changes of the EFG vector may have a weak incidence on the $\UU$
distributions while they may produce significant changes of $\eta_Q$ which is a ratio formed from them. 
This explains the higher sensitivity of the $\eta_Q$ distribution.
To summarize, the distributions of the $U_1$ and $U_{k,k>1}$
of the NBO are largely defined by local structural parameters
(i.e., the number of covalent bonds shared with the phosphorus) whereas
the distribution of the asymmetry parameter $\eta_Q$ is influenced by
second or more remote coordination spheres.

For BO sites, the distributions of the $U_i$'s can be reproduced 
by a two-charge model, where both charges mimic the anisotropy in the electronic structure 
produced by two covalent bonds. The resulting distributions are strongly dependent 
on the angle between the two charges and the centre of the sphere 
(i.e. angle \ce{POP}). This strong dependence of the EFG distribution on
the POP angle is consitent with a previous work of Clark \textit{et al.}~\cite{PRB.70.064202} 
, revisited recently in a MD+DFT-GIPAW approach~\cite{JPCC.113.7917},
who established a correlation between the quadrupolar parameters and the 
\ce{Si-O-Si} angle of \isoO~BO, in silicate glasses. 
For the present two-charge model, 
only three parameters are needed to define the EFG tensor
(five in the general case), correlating the principal components of the tensor ($V_{zz}$ and $\eta_Q$). 
The observed 
EFG distributions can then be reduced to the distribution of two point 
charges with an effective angle. 
The distribution of the BO $U_1$ and $U_{k,k>1}$ are well accounted for by the EFG due to two 
charges with an angle deduced from the MD and DFT calculations. As above, the distributions 
$P(U_k)$ are fully consistent with the requirements of statistical isotropy (section~\ref{subsec:def}). 
Distributions $P(U_{k,k>1})$ due to the effect of two charges making an angle $2\theta$ differ just 
a little from those obtained from uniform distributions of the angle in  domains $[2\theta_1,2\theta_2]$ 
centered on $2\theta$ with a width as large as 16\degre. These angle distributions leave  essentially 
unchanged the distribution $P(U_1)$ (figure~\ref{fig:Ui_EFG} bottom, see in addition~\ref{subsec:append_Uniformrot} 
and figure~\ref{fig:U1_Uk_theo_13}) except for the sharp peak which becomes 
rounded  off. The ability of the $P(U_k)$ distributions to evidence fine effects is again less than it is for 
the $\eta_Q$ distribution.

Finally, the analysis of the ECM distribution for the calculated 
\isoO~EFG parameters shows some limitations to perfectly reproduce 
the $\eta_Q$ distribution. This is partially due to the 
simplified way of modeling  the local contribution to the full EFG tensor by a single fixed tensor (Eq.~\ref{eq:czjzek_etendue}).
Indeed, by simply considering a narrow gaussian distribution of $\eta_Q(0)$,
it is possible to reproduce the observed  
$\eta_Q$ distribution (left top dash-line Figure~\ref{fig:BO_NBO_fit}). 
This additional distribution of the anisotropic contribution could be
interpreted as a small structural distribution of the local geometrical
parameters (distances and angles).

In summary, the $P(U_k)$ are fully consistent with the statistical isotropy of the model glass.  
The modelling of the EFG tensors in terms of the contribution of one or of two charges  
with an additional Czjzek contribution from the background corresponds to the ECM (Eq.~\ref{eq:czjzek_etendue}) 
provided that the local frame of reference is chosen to be the one associated with the 
single charge or the doublet of charges. However, the previous ECM distributions deduced 
from  the MD+DFT calculated EFG parameters of~\isoO~atoms do not account perfectly for their $\eta_Q$ distributions.
This is partially due to a simplified modelling of the local contribution to the 
total ECM (section~\ref{sec:ecm}). As emphasized by Le Ca\"er \textit{et al.} in the conclusion of 
Ref.~\onlinecite{JPCM.22.065402}, a first step to make the ECM more realistic is to distribute the local 
contribution on a physical basis sound. 
Here, a narrow gaussian distribution of $\eta_Q(0)$ is indeed shown to suffice to account for fine details of 
the $\eta_Q$ distribution calculated from the glass model as shown in figure~\ref{fig:BO_NBO_fit} (dashed line, top left subfigure).

\begin{figure}[!t]
\centering
\includegraphics[width=0.45\textwidth]{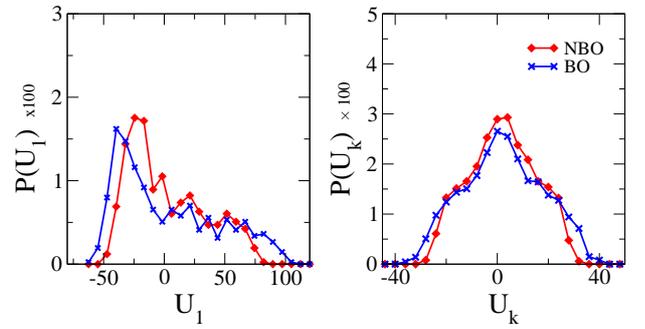}
\caption{(Color on-line) Distributions of $U_1$ and of $U_{k,k>1}$ for the CSA 
tensor calculated for \isoO~NBO (red) and ~\isoO~BO (blue) from the MD configurations.
\label{fig:csa_symetric}}
\end{figure}

\begin{figure*}[!t]
\centering
\includegraphics[width=0.8\textwidth]{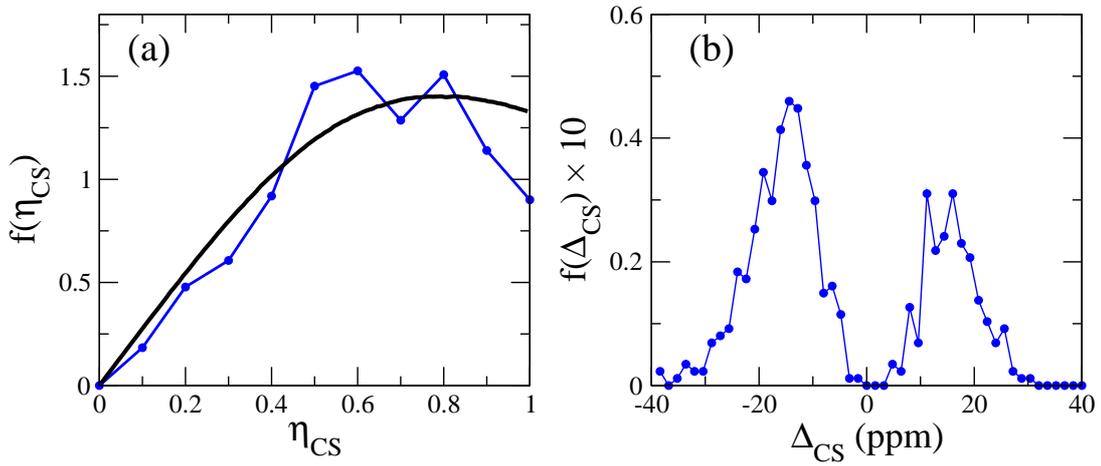}
\caption{$\Delta_{CS}$ and  $\eta_{CS}$ distributions of \isoNa~of the MD
configurations. These two distributions are similar in shape to those of
the related Czjzek EFG distributions (see figure 1). The two parameters $\Delta_{CS}$ 
and $\eta_{CS}$ play respectively the roles of $V_{zz}$  et $\eta_{Q}$.
\label{fig:23Na_CSA}}
\end{figure*}

\subsection{The case of sodium (\isotope[23]{Na})}

The ECM analysis of \isoNa~sites 
reveals a trend of the EFG tensor of these nuclei to be distributed according to a Czjzek model.
This observation was expected when considering the properties 
of a distributed point charge model as presented previously. 
Indeed, a structural analysis of the MD model reveals 
that sodium atoms have a distributed coordination number which ranges between 4 and 8, with more
than 80\% of them being 5- and 6-folded. 
For such coordinations, all the EFG components of the discrete 
charge model are normally distributed. Moreover, Figure~\ref{fig:coord_etaQ} 
presents the distribution of 
$\eta_Q$ for two sodium coordination numbers, $N=$ 5 and 6, from our 
MD configurations. Both $\eta_Q$ distributions show the general 
features of the GIM. Other coordination numbers ($N=$ 4 and 7) 
also show the same trend despite the small sample size (\textit{ca}. 
50 sites for each coordination). More generally, these results confirm 
the statement that the GIM is valid when the coordination number is at 
least equal to 4.~\cite{JMR.192.244,JPCM.10.10715}

		      \section{Towards ECM analysis of the CSA tensor}
\label{sec:CSA}

Both the Czjzek model and the extended Czjzek model, which are based on quite general assumptions, 
are not restricted solely to the analysis of EFG tensors. Physical properties of statistically 
isotropic disordered solids which are represented in 3D space by symmetric second-rank tensors 
can be analysed in a way similar to the one performed in the present paper 
provided that the tensor elements are sums of various contributions to which the central limit 
theorem may apply. These physical properties may be measured or/and calculated from structural 
models. This is for instance the case of the deviatoric part of the atomic level stress tensor 
(ALS) in metallic glasses~\cite{srolovitz1981local,egami2006formation} which was analyzed with 
a GIM in Ref.~\onlinecite{le1984amorphous}. 
The trace of the stress tensor is the local pressure whose distribution is approximately Gaussian~\cite{srolovitz1981local}. 
Similarly, the extended Czjzek model was applied to demagnetizing tensors to account for 
superparamagnetic resonance spectra of ferromagnetic particles in a diamagnetic matrix~\cite{kliava1999size}.

The Chemical Shielding Anisotropy (CSA) tensor is worth being analyzed with 
the previous methods. 
In contrast to the EFG tensor, the CSA tensor has an isotropic part (non-zero trace) and 
is then completely determined by three parameters instead of two ($V_{zz}$ and $\eta_Q$) for the EFG tensor. 
Within the Haeberlen convention,
~\cite{B-Haeberlen} we define the isotropic chemical shielding
($\sigma_{iso}$), the reduced anisotropy ($\Delta_{CS}$) and the asymmetry 
parameter $\eta_{CS}$ from the three principal components of the CSA tensor  ($\sigma_{ii},\ i=1,2,3$). 
\begin{eqnarray}
\sigma_{iso}&=&\frac{1}{3}(\sigma_{11}+\sigma_{22}+\sigma_{33})\\  
\Delta_{CS}&=&\sigma_{33}-\sigma_{iso}\\
\eta_{CS}&=&\frac{\sigma_{22}-\sigma_{11}}{\Delta_{CS}}
\end{eqnarray}
where the components $\sigma_{ii}$ are sorted in the following way
$|\sigma_{33} - \sigma_{iso}| \geqslant |\sigma_{11} - \sigma_{iso}| \geqslant |\sigma_{22} - \sigma_{iso}|$.

In the general case, the CSA tensor is completely characterized by six components. 
Five of them are defined by the five real $U_i$ components given by Eq.~\ref{eq:U_i}, 
substituting $v_{\alpha\beta}$ by $\sigma_{\alpha\beta}$. The additional component, 
denoted as $U_0$, corresponding to the isotropic parameter of the CSA interaction, 
is expressed by the following relation: 
\begin{equation}
\label{eq:U0_CSA}
U_0=-(1/\sqrt 3)[ \sigma_{11} +\sigma_{22} +\sigma_{33} ]\\
\end{equation}
However, to preserve the conditions imposed by statistical isotropy, the $U_1$ component should be now 
defined by 
\begin{equation}
\label{eq:U1_CSA}
U_1= (1/2)[ \sigma_{33} - \sigma_{iso} ]\\
\end{equation}

To illustrate the similarity between the CSA and EFG distributions, 
we analyzed the distribution of $U_{i}$ for the CSA tensor of oxygen sites. 
The data were obtained from the DFT-GIPAW calculation of the \ce{NaPO3} 
configurations generated by MD. 
First, as expected, the distribution of $U_0$ is well described by a gaussian, centered 
around the chemical shielding experimentally observed for both 
NBO and BO (not shown).
Figure~\ref{fig:csa_symetric} shows 
the $U_i$ distributions for the two oxygen environments, namely BO and NBO. 
 The distributions of $U_1$, 
for both oxygens, are asymmetric,  as those observed for the EFG. 
The distributions of the remaining components have a symmetric shape 
which differs from a Gaussian. The distributions of figure~\ref{fig:csa_symetric} show clear 
similarities with those of figure~\ref{fig:Ui_EFG}.
These results lead to 
the conclusion that the~\isoO~CSA distributions need more than the GIM to be described properly.
Therefore, the ECM 
analysis can be used to characterize the CSA distribution, revealing 
at the same time, the local structural information that is encoded into the 
anisotropic part of this interaction.  

Figure~\ref{fig:23Na_CSA} presents the distributions of $\eta_{CS}$ and $\Delta_{CS}$ parameters
for sodium nuclei in our MD structural model. A 
Czjzek-like feature is easily identified for the $\eta_{CS}$ distribution, while the $\Delta_{CS}$
 parameter presents an almost symmetrical distribution about zero, 
and is close to the shape of the $V_{zz}$ distribution in the Czjzek model. 
Both results indicate clearly that the distribution of the CSA tensor obeys the two general assumptions 
which lead to the GIM.

                                             \section{Conclusion}

We applied a general multi-approach whose aim is to analyse and 
to model the NMR tensor distribution in disordered systems.
This approach makes use of MD simulations and DFT calculations of
NMR properties (i.e Electric Field Gradient (EFG) and Chemical 
Shift Anisotropy (CSA)) to provide both structural and NMR information.
 More important, this combined approach 
yields the full distributions of the considered tensors which cannot be measured experimentally in general.
The distributions of the NMR interaction tensors are analysed through 
two different models: the Gaussian Isotropic Model (GIM) also known 
as the Czjzek model and an extension of it, the Extended Czjzek Model (ECM).  
We propose in particular a simple procedure to extract the main 
parameters of these models from a given tensor distribution.

We applied this procedure to a simple binary glass: 
the sodium metaphosphate (\ce{NaPO3}). Using the strong sensitivity 
of NMR to the structural atomic arrangement, we validated
the structural models generated by MD comparing the theoretical 
NMR response to high-resolved Solid-State NMR experiments.
The ECM was used to analyse the calculated distributions of 
the EFG of the two quadrupolar nuclei  present in  this system 
( i.e \isoNa~and \isoO ). The ECM analysis reveals the vailidity of the GIM in case 
of sodium, showing its broad applicability.
As discussed
in previous studies, the universal nature of such distributions (central limit theorem)
makes the extraction of structural information very difficult. 
In case of oxygen, the tensor 
distributions are shown to be dominated by a large local contribution.
A simple additional analysis based on discrete charge distributions showed indeed that 
simple structural information  might be extracted from the distribution of the components $U_i$ of the investigated tensors

Finally, from the simple observations made about the CSA tensor 	
distributions of~\isoO~and~\isoNa~nuclei, it is clear 
that CSA and EFG tensors can be analysed with similar tools.
In other words, both GIM and ECM analyses might be useful and relevant to discuss  
experimental and theoretical observations of CSA distributions. 

      \acknowledgments

Some of the numerical results presented in this paper were carried 
out using the regional computational cluster supported by Universit\'e Lille 1, 
CPER Nord-Pas-de-Calais/FEDER, France Grille and CNRS. We highly 
appreciate and thank the technical staff of the CRI-Lille 1 center for their strong and helpful support. 
IDRIS is also acknowledged for CPU allocation under project number x20080911849. 
FV was supported by the Minist\`ere de 
l'\'Education Nationale de l'Enseignement Sup\'erieure et de la Recherche. 
The FEDER, R\'egion Nord Pas-de-Calais, Minist\`ere de l'\'Education Nationale 
de l'Enseignement Sup\'erieur et de la Recherche, CNRS, and USTL are 
acknowledged for financial support. FV would acknowledge previous support
from the Stichting voor Fundamenteel Onderzoek der Materie (FOM).

							\appendix

\section{Additional remarks on the ECM }
\label{sec:appendix1}

All the ECM tensors $\VV(\epsilon)$ are expressed in a local frame of reference which is chosen,
without loss of generality, as the one in which $\VV_0$ is diagonal and changes from atom
to atom. 
By contrast, the distribution of the EFG tensor $\VV'(\epsilon)$ 
(presented in \ref{sec:appendix2})
of all ``sites'' belonging to the same family is calculated in a fixed global frame
of reference identical for all atoms. 
It differs thus from the distribution of the tensor $\VV(\epsilon)$ (Eq.~\ref{eq:czjzek_etendue}).
The spatial extent of a cluster around the considered
atomic probe is expected to depend on the investigated solid and on the physical
origins of the EFG (ex: close neighborhood for covalent glasses). The statistical
isotropy of this family would mean that the previous clusters have an overall random
orientation in the considered solid. This general property holds for any geometrical
characteristics of these clusters. The only general a priori knowledge about the distribution of $\VV'(\epsilon)$
at the cluster centers is that it depends on the invariants of the EFG tensor when statistical isotropy
holds (see equations 7,8 and 10 of Ref.~\onlinecite{PRB.23.2513} ). In addition, the associated distributions
$P(U'_k)$ fulfill the conditions described in section~\ref{subsec:def}. To obtain the isotropic distribution
of $\VV'(\epsilon)$ from Eq.~\ref{eq:czjzek_etendue} , it would suffice to rotate
the local frame of reference, defined from $\VV_0$, uniformly in all directions.
The principal value distribution of  $\VV'(\epsilon)$ is identical with the
principal value distribution of $\VV(\epsilon)$ as further 
discussed in~\ref{sec:appendix2}.
Eq.~\ref{eq:czjzek_etendue} suffices thus to derive the required information 
about the $\eta_Q$ and $V_{zz}$ distributions of the associated
statistically isotropic EFG. A consequence of this choice is however that the distribution $P(\UU')$,
of $\UU'(\epsilon)$ the vector associated with $\VV'(\epsilon)$ (Eq.~\ref{eq:U_i}),
cannot be derived directly from Eq.~\ref{eq:czjzek_etendue}. 

\section{Rotation of the ECM tensor}
\label{sec:appendix2}

Consider a particular tensor $\VV(\epsilon)$ obtained by choosing at random 
some $\VV_{\mathrm{GIM}}$ tensor in Eq.~\ref{eq:czjzek_etendue}, 
$\VV(\epsilon)=\VV_0+\rho(\epsilon)\VV_{\mathrm{GIM}}$, where $\VV_0$ is diagonal. 
The considered symmetric second-rank tensor is transformed into a diagonal 
tensor $\VV_{\mathrm D}(\epsilon)$ by some rotation whose associated (3x3) matrix is $\mathrm{H}_0$, 
in matrix notation:
\begin{equation}
\label{eq:diagtens}
\VV_{\mathrm D}(\epsilon) = \mathrm{H}_0 \VV(\epsilon) \mathrm{H}_0^\intercal
\end{equation}

and  thus, $\VV_{\mathrm D}(\epsilon) = \mathrm{H}_0^\intercal \VV(\epsilon) \mathrm{H}_0$ , 
where $\mathrm{H}_0^\intercal$ is the transpose of $\mathrm{H}_0$ 
($\mathrm{H}_0\mathrm{H}^\intercal=\mathrm{H}_0^\intercal\mathrm{H}_0=\mathrm{I}$). 
The  tensor $\VV(\epsilon)$ is transformed by a rotation $\mathrm{H}_1$ into a tensor $\VV'(\epsilon)$:
\begin{equation}
\label{eq:rottens}
\VV'(\epsilon) = \mathrm{H}_1 \VV(\epsilon) \mathrm{H}_1^\intercal = \mathrm{H}_1 \mathrm{H}_0^\intercal \VV_{\mathrm D}(\epsilon) \mathrm{H}_0 \mathrm{H}_1^\intercal
\end{equation}

or equivalently: 
\begin{equation}
\label{eq:rottens2}
\VV'(\epsilon) = \mathrm{H}_2 \VV_{\mathrm D}(\epsilon) \mathrm{H}_2^\intercal 
\end{equation}
 
where $\mathrm{H}_2=\mathrm{H}_1 \mathrm{H}_0^\intercal$ is a rotation matrix. 
The distribution of $\mathrm{H}_1$ must be taken as uniform over the special orthogonal group SO(3)
to obtain a global statistical isotropy of the EFG tensor $\VV'(\epsilon).$
The standard definition of uniformity requires the distribution of $\mathrm{H}_1$ to remain 
unchanged when composed with any arbitrary rotation (Haar measure). 
Therefore, the distribution of $\mathrm{H}_2$ is uniform too and the principal values of the tensors $\VV'(\epsilon)$ and 
$\VV(\epsilon)$ coincide with the diagonal elements of $\VV_{\mathrm D}(\epsilon)$.  This line of reasoning can 
be applied to any tensor formed as above from Eq.~\ref{eq:czjzek_etendue} . Thus, by construction, the global 
distribution of the EFG tensor $\VV'(\epsilon)$ is statistically isotropic and its principal value 
distribution is identical with that of $\VV(\epsilon)$. Therefore, the knowledge of the a priori 
complicated distribution of $\VV'(\epsilon)$ is useless when only principal values matter.  Equation~\ref{eq:czjzek_etendue} 
suffices to provide the required information about the $\eta_Q$ and $V_{zz}$ distributions of the 
associated statistically isotropic EFG tensor $\VV'(\epsilon)$. However, by this choice, the 
distribution $P(\UU'(\epsilon))$ 
cannot be derived without additional calculations based on Eq.~\ref{eq:rottens2}. Simplified examples 
are given below to illustrate the need of more than a single family of sites to 
account for the EFG distribution of some disordered solids described by the ECM.

\begin{figure}[!t]
\centering
\includegraphics[width=0.5\textwidth]{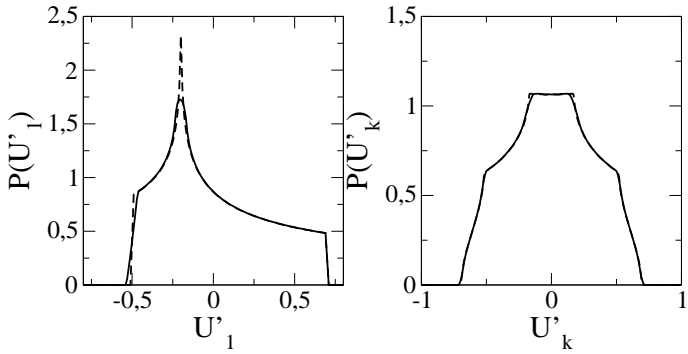}
\caption{Distribution of $U'_1$ and of $U'_{k,k>1}$ obtained with $\beta_0=$ 0.7 either with $\eta_Q(0)=3/7$ (dotted lines) or with $\eta_Q(0)$
uniformly distributed between 0.33 and 0.53 (solid lines). The distribution of $P(U'_1)$ with a fixed $\eta_Q(0)$
is calculated from equations~\ref{eq:pz_1} and~\ref{eq:pz_2} while all the others distributions are calculated by Monte-Carlo simulations.
\label{fig:U1_Uk_theo_13}}
\end{figure}

\begin{figure}[!t]
\centering
\includegraphics[width=0.5\textwidth]{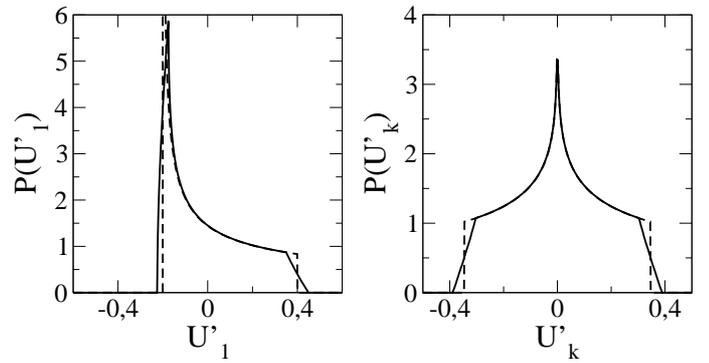}
\caption{Distributions of $U'_1$ and of $U'_{k,k>1}$ obtained from relations~\ref{eq:ex1_onecharge_U1} and~\ref{eq:ex1_onecharge_Uk}.
The distributions are calculated for (dashed line) a fixed $\beta_0=$ 0.4 and (full line) a $\beta_0$ parameter uniformly distributed between 0.35 and 0.45.
\label{fig:U1_Uk_theo_12}}
\end{figure}

\subsection{Uniform rotation}
\label{subsec:append_Uniformrot}
We consider the fixed diagonal part $\VV_0$ of the ECM model (Eq.~\ref{eq:czjzek_etendue}) which is 
expressed in its system of principal axes. The associated vector $\UU_0$  has therefore only 
two non-zero components, namely $U_1(0)=V_{zz}(0)/2$ and $U_5(0)=\eta_Q(0)V_{zz}(0)/2\sqrt{3}$ 
When the principal axis system is rotated uniformly in all directions, the tensor $\VV_0$
is transformed into a tensor $\VV'_0$ whose elements are consequently distributed. By construction, the
distribution of $\VV'_0$ is then statistically isotropic and the components of the associated vector $\UU'_0$ have
marginal distributions which fulfill the conditions described in section~\ref{subsec:def}. 
The distributions $P(U'_k(0))$ can be obtained either in closed form 
for $k=1$ and for any $k$ for $\eta_Q(0)=$ 0 or by numerical simulations 
for any $k>1$ as soon as $V_{zz}(0)$ and $\eta_Q(0)$ are known. 
These two parameters are obtained in the examples discussed below from simple  physical models.
This appendix aims to present the distribution of the first component $U'_1(0)$ of $\UU'_0$. 

After a rotation of the previous frame of reference by the Euler angles $\alpha,\beta,\gamma$, 
with ($0\leq \alpha,\gamma<2\pi,0\leq\beta<\pi$), $U'_1(0)$ is directly calculated from the expressions 
given in~\ref{sec:appendix2} of Ref.~\onlinecite{JPCM.10.10715} to be: 
\begin{equation}
\label{eq:appendiXD}
U'_1(0)=\frac{\beta_0}{2}(3\cos^2(\beta)-1+\eta_Q(0)\sin^2(\beta)\cos(2\alpha)) \\ 
\end{equation}
with 
\begin{equation}
-\rho_0<U'_1(0)<\beta_0
\end{equation}
where $\beta_0$ is some positive scale factor and $\rho_0=\beta_0(1+\eta_Q(0))/2$. 
To derive the sought-after distribution, it suffices therefore to consider a random rotation  
($0\leq \alpha<\pi/2,0\leq\beta<\pi/2$) with a weight $(2/\pi)\sin\beta \mathrm{d}\alpha \mathrm{d}\beta$.
Focusing on the angular part of Eq.~\ref{eq:appendiXD} we obtain two integrals which depends on
$\cos(2\alpha)$. The first integral is obtained when $U'_1(0)< \beta_0(\eta_Q(0)-1)/2$ and the second when 
$U'_1(0) > \beta_0(\eta_Q(0)-1)/2$. The final distribution reads ($z=U'_1(0)$)
\begin{align}
\label{eq:pz_1}
\begin{cases}
P(z)=\mathrm{_{2}F}_1\big(\frac{1}{2},\frac{1}{2};1;A\big) \Big(4\beta_0\eta_Q(0)(\beta_0-z)\Big)^{-1/2}\\
-\rho_0<z<\rho_0-\beta_0\\
\end{cases}
\end{align}

\begin{align}
\label{eq:pz_2}
²\begin{cases}
P(z)=\mathrm{_{2}F}_1 \big ( \frac{1}{2},\frac{1}{2};1; \frac{1}{A} \big) \Big(2\beta_0\big(3-\eta_Q(0)\big)\big(\rho_0+z\big)\Big)^{-1/2}\\
\rho_0-\beta_0< z < \beta_0\\
\end{cases}
\end{align}
with
\begin{equation}
A=\Big(\big(3-\eta_Q(0)\big)\big(\rho_0+z\big)\Big) \Big(2\eta_Q(0)\big(\beta_0-z\big)\Big)^{-1}
\end{equation}
where $\mathrm{_{2}F}_1(,;;)$ is a hypergeometric function. The previous 
probability density function is discontinuous at $U'_1(0)=\rho_0-\beta_0=\beta_0(\eta_Q(0)-1)/2$. 

As a first example, we consider the case of two identical charges $q_1$ and $q_2$ are located at equal distances from an origin $O$ 
at which the EFG is observed, with an angle $\widehat{q_1Oq_2}=2\theta$. The
asymmetry parameter is $\eta_Q(0)=3\cos^2\theta/(2-3\cos^2\theta)$ when $\theta=\mathrm{arcos}(1/\sqrt{3})<\theta<\pi/2$.
If $\theta$ is taken equal to $\mathrm{arcos}(1/\sqrt{5})\sim$63.4\degre, the angle 
between the two charges is then 126.8\degre and $\eta_Q(0)=3/7\sim$0.4286. 
The distribution $P(U'_1(0))$, which is calculated in that way from Eqs.~\ref{eq:pz_1} and~\ref{eq:pz_2}
for $\beta_0=$ 0.70 (Fig.~\ref{fig:U1_Uk_theo_13}), reproduces  then the essential 
characteristics of the binned distribution $P(U'_1)$ shown in figure~\ref{fig:Ui_EFG} 
for the case of two charges. As explained in section~\ref{subsec:STRUCTanalysis_17O}, 
contributions from fluctuations of $\VV_0$ and from the Czjzek background must 
however be added to account for the details of $P(U'_1)$.

When $\eta_Q=$ 1, the components $U_1$ and $U_5$ cannot be distinguished, except for a scale factor,
because $V_{xx}=$ 0 and $V_{yy}=-V_{zz}$. Therefore, the distribution $P(U'_1(0))$ must become symmetric as
is $P(U'_5(0))$ when statistically isotropy holds. Indeed,  $P(U'_1(0))$ reduces to 
\begin{equation}
P(U'_1(0))=\frac{\mathrm{_{2}F}_1\Big( \frac{1}{2} , \frac{1}{2} ; 1 ; \frac{\beta_0-|U'_1(0)|}{\beta_0+|U'_1(0)|} \Big)} { 2 \sqrt{ \beta_0 (\beta_0 + |U'_1(0)|}}  
\end{equation}
for
\begin{equation}
-\beta_0 <  U'_1(0) < \beta_0 \;\;\; \mathrm{as}\;\;\; \rho_0=\beta_0
\end{equation}
The latter distribution is similar to the distribution $P(U'_k)$ calculated for a single charge.
Using the statistical invariance by rotation, the variance of $U'_1(0)$ is readily obtained to be :
\begin{equation}
\langle U'_1(0)^2\rangle = \frac{\beta_0^2}{5}\Bigg(1+\frac{\eta_Q(0)^2}{3}\Bigg)
\end{equation}

\subsection{The case of a single charge}
\label{subsec:append_onecharge}

As a second example, we consider the EFG at the nuclei of a given
isotope which is determined by a single neighboring point charge
and by more remote atomic shells whose total contribution is described
by a Czjzek tensor. The $\VV_0$ contribution to Eq.~\ref{eq:czjzek_etendue} is then due to
this single charge. If the charge does not fluctuate in value and
in distance from the considered atom, the total EFG is described
by a single ECM  (Eq.~\ref{eq:czjzek_etendue}) with $\eta_Q(0)=$ 0. If the charge fluctuates
in value or(/and) in distance, then the EFG must be described by
a distribution of ECM, all with $\eta_Q(0)=$ 0. 
We focus below on the distribution of the sole local part. The effect of the addition of a
small Czjzek noise  is just a smoothing of the distributions shown in figure~\ref{fig:U1_Uk_theo_12}.

The marginal distributions of the components of ($\UU'_0=(U'_1,\ldots,U'_5)$), are obtained
from Eq.~\ref{eq:allU} with $n=$ 1 for a random distribution of $\theta$ and $\phi$
(with $P(\theta,\phi)=\frac{1}{4}sin(\theta)d\theta d\phi$). The distribution
of $P(U'_1)$ is asymmetrical as expected from the conditions given at the end of section~\ref{subsec:def} and has
a simple closed-form:
\begin{equation}
\label{eq:ex1_onecharge_U1}
P(U'_1)= (3\beta_0^2 + 6\beta_0 U'_1)^{-1/2} \;\;\;\;   \Big(-\frac{\beta_0}{2} < U'_1 < \beta_0 \Big )
\end{equation}

where the proportionality constant $\beta_0$ is chosen here to be positive and is related to
the characteristic parameters (i.e charge and distance) of the problem.
By contrast, the distributions  $P(U'_{k,k>1})$ are all identical and symmetric:

\begin{equation}
\label{eq:ex1_onecharge_Uk}
P(U'_{\mathrm k})= \frac{\mathrm{_{2}F}_1\Big( \frac{1}{2} , \frac{1}{2} ; 1 ; \frac{\alpha-|U'_k|}{\alpha+|U'_k|} \Big)} { 2 \sqrt{ \alpha (\alpha + |U'_k|)}}  \;\;\;\;  -\alpha <  U'_k < \alpha
\end{equation}

where  $\alpha=\beta_0\sqrt{3}/2$ and $\mathrm{_{2}F}_1(,;;)$ is a hypergeometric function.
In figure~\ref{fig:U1_Uk_theo_12}, we represent two examples of distributions obtained from Eqs.~\ref{eq:ex1_onecharge_U1} and~\ref{eq:ex1_onecharge_Uk}: with and without a uniform distribution of $\beta_0$ where $\beta_0$ is chosen to reproduce the characteristic distribution observed in our structural model (see the binned distribution, figure~\ref{fig:Ui_EFG} top right, solid line).

\subsection{A perturbed diamond lattice}
\label{subsec:perturbed_diamond}

A third example deals with a diamond lattice whose atomic positions are very 
slightly shifted with random displacements derived from an isotropic Gaussian 
distribution. For simplicity, we assume that the EFG is due to point charges 
located at the lattice sites.  In addition, the EFG at any site of this perturbed 
diamond lattice is assumed to be essentially due to its four first neighbors while 
more remote shells contribute to a Czjzek background. We focus on the distribution 
of $\eta_Q$ obtained from $\VV(\epsilon)$ (Eq.~\ref{eq:czjzek_etendue}). 
Indeed, although the principal values of 
$\VV(\epsilon)$ have very small magnitudes, they suffice to change significantly $\eta_Q$ as 
it is a ratio. An expansion of the EFG tensor in terms of the Gaussian atomic 
displacements gives a zero-order tensor $\VV_0$ which comes from the tetrahedral 
coordination of a site and a first-order EFG tensor whose elements are centered Gaussians.   
This expansion yields Eq.~\ref{eq:czjzek_etendue} as the first-order tensor 
is a Czjzek tensor by itself. When added to the background tensor, the sum is 
still a Czjzek tensor because the sum of two independent Czjzek tensors with 
parameters $\sigma_1$ and  $\sigma_2$ is a Czjzek tensor whose parameter is $\sigma=\sqrt{\sigma_1^2 +\sigma_2^2}$. If all atoms are 
identical, $\VV_0=$ 0 and the EFG distribution is then a Czjzek distribution 
If two different atomic species A and B, with different 
charges $q_A$ and $q_B$ ($\Delta q=|qA-qB|$), are distributed at random on the lattice 
sites with a composition \ce{A_xB_{1-x}}, without any change of the atomic positions, 
then the distribution of $\eta_Q$ at the A nuclei is a mixture of three distributions:
\begin{itemize}
\item[(i)] a Czjzek distribution which comes from A atoms, denoted \ce{A[A_4]} and \ce{A[B_4]},  
which are  surrounded respectively by 4A atoms and by 4B atoms, with a weight ($x^4+(1-x)^4$) ($V_{zz}(0)=$ 0). 
\item[(ii)] an ECM distribution which originates from \ce{A[A_3B]} and \ce{A[AB_3]} atoms, with a weight ($4x^3(1-x)+4x(1-x)^3$ ) and $\eta_Q(0)=$ 0 ($V_{zz}(0)\propto2\Delta q$).
\item[(iii)] an ECM distribution which arises from \ce{A[A_2B_2]}  atoms, with a weight $6x^2(1-x)^2$ and $\eta_Q(0)=$ 1, ($V_{zz}(0)\propto\Delta q$).
\end{itemize}
In (ii) and (iii), the tensor $\VV_0$ and the values of $\eta_Q(0)$ reflect the symmetries 
of the underlying tetrahedral shell. 
The previous example, although oversimplified, sketches however situations which 
are encountered in real semiconductors. Indeed, a recent \isoAs~and \isoGa~solid 
state NMR study of \ce{Al_xGa_{1-x}As} thin films combines on the one hand experimental 
results and on the other structural modeling of disorder in the \ce{Ga} and \ce{Al} positions 
together with first-principles DFT calculations to obtain, among others, the EFG 
distributions at the tetrahedrally coordinated As sites (Ref.~\onlinecite{PCCP.12.11517}). 
Knijn \textit{et al.} found that the EFG distributions at \ce{As[Al_4]}, \ce{As[Ga_4]} sites are 
accurately described by the Czjzek distribution. In addition, they showed that  the 
EFG distributions at  the \ce{As} sites \ce{As[Al_1Ga_3]}, \ce{As[Al_3Ga_1]} and \ce{As[Al_2Ga_2]} are accurately described by 
extended Czjek distributions whose $\eta_Q(0)$ are respectively 0, 0, 1  as above 
for symmetry reasons (Figure 13 of Ref.~\onlinecite{PCCP.12.11517} ).

\section*{References}

\bibliographystyle{apsrev}
\bibliography{refecm}

\end{document}